\PassOptionsToPackage{dvipsnames}{xcolor}

\documentclass[acmsmall,authorversion,nonacm]{acmart}

\AtBeginDocument{%
	}

\setcopyright{acmcopyright}
\copyrightyear{2024}
\acmYear{2024}
\acmDOI{XXXXXXX.XXXXXXX}

\usepackage{enumitem}
\usepackage{float}

\usepackage{multirow}
\usepackage{makecell}

\usepackage{csquotes}
\usepackage{amsmath}
\usepackage{bm}
\robustify\bfseries

\usepackage{siunitx}
\usepackage[position=top]{subfig}
\usepackage{graphicx}

\usepackage{pifont}
\newcolumntype{P}[1]{>{\centering\arraybackslash}m{#1}}
\newcolumntype{R}[1]{>{\raggedright\arraybackslash}m{#1}}
\newcolumntype{L}[1]{>{\raggedleft\arraybackslash}m{#1}}

\usepackage{xspace}
\newcommand\ie{i.\,e.\xspace}
\newcommand\eg{e.\,g.\xspace}

\newcommand\abs[1]{| #1 |}

\usepackage{url}

\makeatletter
\renewcommand{\fps@figure}{htb}         
\renewcommand{\fps@table}{htb}         
\makeatother 

\usepackage[ruled]{algorithm2e}
\usepackage{cleveref}

\newcommand{\FINAL}[1]{}

\begin{document}

\title{The Virality of Hate Speech on Social Media}

\author{Abdurahman Maarouf}
\affiliation{%
  \institution{LMU Munich \& Munich Center for Machine Learning}
  \country{Germany}}
\email{a.maarouf@lmu.de}

\author{Nicolas Pröllochs}
\affiliation{%
  \institution{JLU Giessen}
  \country{Germany}}
\email{nicolas.proellochs@wi.jlug.de}

\author{Stefan Feuerriegel}
\affiliation{%
  \institution{LMU Munich \& Munich Center for Machine Learning}
  \country{Germany}}
\email{feuerriegel@lmu.de}



\begin{abstract}
Online hate speech is responsible for violent attacks such as, e.g., the Pittsburgh synagogue shooting in 2018, thereby posing a significant threat to vulnerable groups and society in general. However, little is known about what makes hate speech on social media go viral. In this paper, we collect $N = \num[group-separator={,},group-minimum-digits=1]{25219}$ cascades with $ \num[group-separator={,},group-minimum-digits=1]{65946}$ retweets from X (formerly known as Twitter) and classify them as hateful vs. normal. Using a generalized linear regression, we then estimate differences in the spread of hateful vs. normal content based on author and content variables. We thereby identify important determinants that explain differences in the spreading of hateful vs. normal content. For example, hateful content authored by verified users is disproportionally more likely to go viral than hateful content from non-verified ones: hateful content from a verified user (as opposed to normal content) has a \num{3.5} times larger cascade size, a \num{3.2} times longer cascade lifetime, and a \num{1.2} times larger structural virality. Altogether, we offer novel insights into the virality of hate speech on social media.
\end{abstract}

%



\received{July 2023}
\received[revised]{October 2023}
\received[accepted]{November 2023}

\maketitle



\section{Introduction} \noindent 

Hate speech on social media poses a widespread problem of societal significance. In 2021, around 40\,\% of the U.S. society has personally experienced online hate speech \cite{Vogels.2021}. Hate speech is known to have a negative impact on the mental as well as physical well-being of online users. In particular, young adults tend to suffer from the psychological consequences \cite{Saha.2019}.  In the context of the Global South, online hate speech often serves as a tool to seed fear towards ethnic and religious minorities, especially in the political discourse \cite{Chandra.2021, Saha.2021, Jafri.2023}. Online hate speech further reinforces hateful attitudes (i.e., radicalization) and motivates violent acts including hate crimes \cite{Muller.2021,Bar.2023}. For example, online hate speech has played a crucial role in the 2018 Pittsburgh synagogue shooting \cite{Roose.2021}, the 2017 Rohingya genocide in Myanmar \cite{Mozur.2021}, and anti-Muslim mob violence in Sri Lanka \cite{Taub.2021}.


While negative consequences of hate speech have been widely documented, our understanding of how hateful content spreads on social media is still emerging. Prior work has studied targets of hate speech, which are characterized by higher social activity, older accounts, and more followers \cite{ElSherief.2018}. Moreover, several studies have analyzed characteristics of hateful users \cite{Ribeiro.2018, Hua.2020, Mathew.2020}. Those have been found to be associated with a higher social media activity, more complex word usage, and a higher density in terms of connections among each other, compared to normal users. Social media cascades by hateful users have typically a larger spread, longer lifetime, and higher virality \cite{Mathew.2019}. However, all of these analyses were conducted at the user level, and \underline{not} at the content level. In particular, no prior work analyzed the differences in (and the determinants for) the spreading dynamics of hateful vs. normal content.


We hypothesize that differences in the spreading of hateful vs. normal online content can be explained by characteristics of both the author and the content. For instance, in terms of author characteristics, one may expect that a user with a verified status disproportionately promotes the spreading of hateful content. The reason is that a verified status lends the author a higher credibility and trustworthiness \cite{Morris.2012}, and, as a result, others may be more willing to share hate speech if it comes from a figure with large social influence. Moreover, societal leaders play an important role in shaping the social norms of other users \cite{Siegel.2020}. Hence, individuals with a special social status (such as a verified status on social media platforms) may be able to let hateful content appear socially acceptable, thus adding to the proliferation of hate speech. In terms of content characteristics, one may expect that, for instance, the use of mentions (i.e., using \textquote{@username} in tweets) affects the spread of hate speech. Mentions make hate speech appear directed against a specific user or entity, which differs from generalized hate that is against a general group of individuals \cite{ElSherief.2018b}. Hence, it is likely that directed vs. generalized hate also have different spreading dynamics. Note that our results later provide empirical evidence in favor of these hypotheses (e.g., we find that a verified status adds to the proliferation of hate speech, while mentions and hashtags hamper the spread).


In this work, we analyze differences in the spreading of hateful vs. normal content on social media. Specifically, we aim to generate an understanding of what makes online hate speech go viral:

\vspace{0.2cm}
\noindent\textbf{RQ: } \emph{What are the determinants for the differences in the spreading dynamics of hateful vs. normal content on online social media?}
\vspace{0.2cm}

\noindent
\textbf{Data: } We collected a comprehensive dataset of $N = \num[group-separator={,},group-minimum-digits=1]{25219}$ retweet cascades from X (formerly known as Twitter). Each cascade was human-labeled into two categories: hateful and normal. Our dataset comprises both the root tweet and all retweets, i.e., the complete retweet cascades. The root tweets in our dataset have received $\num[group-separator={,},group-minimum-digits=1]{65946}$ retweets by $\num[group-separator={,},group-minimum-digits=1]{62051}$ different users. For each root tweet, we collected an extensive set of both author characteristics (i.e., number of followers, number of followees, verified status, tweet volume, and account age) and content characteristics (i.e., media items, hashtags, mentions, and tweet length).


\noindent\textbf{Methods:}\footnote{Code and data for our analysis are available via \url{https://github.com/abdumaa/hatespeech_virality}.} We perform an explanatory regression analysis using generalized linear models (GLMs). Thereby, we aim to identify author and content characteristics that explain the differences in the spread of hateful vs. normal content. To quantify the spread on social media, we analyze the following structural properties: (i)~cascade size, i.e., how many times the root tweet was retweeted and which thus measures the overall exposure; (ii)~cascade lifetime, i.e., how long the cascade lived; and (iii) structural virality, i.e., a measure for the effectiveness of the spreading process \cite{Goel.2016}.

\noindent\textbf{Findings:} To the best of our knowledge, this is the first work identifying what makes hate speech go viral. Using explanatory regression modeling, we yield the following novel findings:
\begin{enumerate}
    \item Cascades with hateful content (as compared to cascades with normal content) grow larger in size, live longer, and are of larger structural virality.
    \item Author characteristics explain differences in the spread of hateful vs. normal content. In particular, cascades with hateful content from verified authors are disproportionally larger in size, longer in lifetime, and higher in structural virality.
    \item Content characteristics explain differences in the spread of hateful vs. normal content. In particular, cascades with hateful content including mentions or hashtags are associated with a smaller size, lifetime, and structural virality.
\end{enumerate}

\section{Related Work} 

In the following, we review different literature streams that are particularly relevant to our work, namely our subject (online hate speech) and our methods (modeling spreading dynamics).




\subsection{Hate Speech on Social Media}


Hate speech has been defined by the United Nations Strategy and Plan of Action as {\textquote{any kind of communication in speech, writing or behaviour, that attacks or uses pejorative or discriminatory language with reference to a person or a group on the basis of who they are, in other words, based on their religion, ethnicity, nationality, race, colour, descent, gender, or other identity factor.}} While the legal definitions vary across countries, the common principle is that hate speech expresses animosity against specific groups, often even calling for violence. 


Authors of hateful content on X have several characteristics that distinguish them from normal authors. For example, hateful content is typically created by authors with a higher tweet activity, a more specific, non-trivial word usage, and a more dense connection among each other, as compared to normal users \cite{Ribeiro.2018, Mathew.2020}. Online hate is also a common concern in the Global South: Prior research on social media in the Global South characterizes hateful users by creating and spreading an illusion of fear, thereby eliciting hate towards minority groups \cite{Saha.2021}. For example, in India, hateful users are particularly active in the political discourse by using symbols and past events (e.g., the COVID-19 pandemic) to spread hate against contesting parties and minorities \citep{Chandra.2021, Jafri.2023}. Moreover, hateful users in the Global South are characterized as having a larger following as well as being more verified, leading to a more polarized and influenced audience \cite{Dash.2022}. Some works focus also on specific types of online hate speech such as hate against politicians \cite{Hua.2020} or religious hate \cite{Albadi.2019,Kursuncu.2019}.


Other works examine the users targeted by hate speech. For example, figures from American politics are more likely to receive hateful replies on X when they are persons of color from the Democratic party, white Republicans, women, and/or authors of content with negative sentiment \cite{Solovev.2022a}. Moreover, users employing moralized language are more likely to receive hate speech \cite{Solovev.2023}. In general, hate speech on X is predominantly addressed against users with an older account, a higher tweet activity, and more followers, as compared to normal users \cite{ElSherief.2018}. 

Note that the above works perform analyses at the \emph{user} level, but not at the \emph{content} level. These works can answer \emph{who} composes/receives hate speech but not how hate \emph{travels}. Hence, the spreading dynamics of hate speech on social media remain unknown.

\subsection{Spreading Dynamics on Social Media} 


Previous research has aimed at a better understanding of the spreading dynamics of social media content and thus why some tweets go viral and others do not. For this, different structural properties of retweet cascades have been analyzed: (1)~cascade size, (2)~cascade lifetime, and (3)~structural virality. Here, \emph{cascade size} refers to how many times a tweet was retweeted and thus measures the overall exposure \cite{Bakshy.2011, Myers.2014, Taxidou.2014, Zang.2017}. The \emph{cascade lifetime} refers to the overall duration the cascade was active \cite{Macskassy.2011, Subbian.2017}. The so-called \emph{structural virality} is a statistical metric to quantify of the trade-off between depth and breadth of the cascade and should thus capture how effectively a tweet propagates \cite{Goel.2016}. The definition of structural virality is related to the Wiener index, and thus measures the average distance between all pairs of nodes in the cascade.


Differences in the structural properties of cascades have been found across different dimensions. These can be grouped into author characteristics and content characteristics. (1)~Author characteristics include, e.g., number of followers, number of followees, tweet volume (i.e., the previous activity on X), account age, and verified status \cite{Suh.2010}. For example, authors with more followers are theorized to have larger \textquote{social influence,} and their tweets should thus reach a larger audience \cite{Stieglitz.2013, Zaman.2014, Vosoughi.2018,Prollochs.2023}. (2)~Content characteristics include, e.g., the use of media items (e.g., images, videos, URLs), mentions, and hashtags \cite{Stieglitz.2013, Goel.2016, Vosoughi.2018}. For example, URLs and hashtags have been found to be positively associated with spreading dynamics \cite{Suh.2010}.

\subsection{The Spread of Online Hate Speech}


There are only a few works that have modeled the spread of online hate speech, yet with a different objective from ours. On the one hand, there are predictive models that aim to forecast certain behavior. For example, some works aim to predict the probability with which hate speech is retweeted \cite{Cheng.2014, Lin.2021}. However, predictive models are only trained on hateful content and, therefore, do not offer explanations on why the spread of hateful vs. normal content differs.


On the other hand, there is one work making quantitative comparisons between hateful and non-hateful authors on Gab \cite{Mathew.2019}. Therein, the authors create a labeled dataset of hateful vs. normal posts and reposts on the platform Gab. The authors report summary statistics and find that cascades by hateful users live longer, are larger in size, and exhibit more structural virality. Moreover, they find that authors of hateful content are more influential, cohesive, and proactive than authors of normal content. However, this work is different from ours in that it compares the spread by hateful vs. normal authors, but \underline{not} by hateful vs. normal content. Hence, the determinants for why hateful content goes viral remain unknown.

\vspace{0.2cm}
\noindent
\textbf{Research gap:} To the best of our knowledge, there is no prior work studying what makes hate speech go viral on online social media. For this reason, we provide the first analysis identifying determinants that explain differences in the spread of hateful vs. normal content on online social media.

\section{Data}

\subsection{Data Collection}


We analyze a comprehensive dataset of hateful and normal tweets on the social media platform X \cite{Founta.2018}. We select X in our study as it represents a platform with high popularity. In 2022, it counted around 436 million monthly active users.\footnote{https://www.statista.com/} Moreover, hate speech is still widespread on X. Estimates suggest that there were around \num{1.13} million accounts on X violating against their policy banning hateful content in the second half of the year 2020, which marks an increase of 77\% compared to the first half of the year.\footnote{https://blog.twitter.com/en\_us/topics/company/2021/an-update-to-the-twitter-transparency-center}  


We use the dataset from \citet{Founta.2018} in which the tweets were human-labeled via crowdsourcing into two categories: hateful and normal. The dataset was created using a boosted sampling procedure, ensuring an unbiased dataset as well as a large number of annotations for the minority class of hateful tweets. The final labels were generated through an iterative process of multiple annotation rounds. In the first rounds, annotators were given brief conceptual definitions, allowing for exploratory annotations. The final definitions and design choices for the annotation process were determined based on the evaluations of these exploratory rounds. As a result, the subsequent and final round benefits from an optimized design choice and more precise definitions, resulting in more accurate annotations. The final label for each tweet is determined based on a majority vote of five crowdsourced annotations. For more details on the dataset, we refer to \citet{Founta.2018}.

We additionally process the dataset as follows. Using the Twitter Historical API, we first retrieve the type of each tweet, that is, whether it is a root tweet, a reply, or a retweet. To achieve this, we extract the field \textquote{referenced\_tweet} (using the \textquote{tweet\_id} provided by \citet{Founta.2018}) which contains the type of the tweet as well as the id of the root tweet. We replace retweets with their corresponding root tweet. As the text of the root tweet and its retweets are identical, we kept the same annotation in case of replacement. In addition, we filter out replies, such that the dataset only consists of root tweets. In the next step, we collect all retweets for each root tweet in order to construct the retweet cascades. Here, we follow the methodological approach in \cite{Vosoughi.2018} to retrieve the underlying retweet paths. Specifically, we first query the Twitter Historical API for all retweets that refer to the ids of the root tweets using the information provided in the field \textquote{referenced\_tweet}. While tweets and retweets have timestamps, the Twitter Historical API does not provide the true retweet path. Instead, all retweets point to the root tweet, which does not reflect the true retweet cascade (i.e., users can also retweet the retweets of other users, not the root tweet). To overcome this, we use the method of \textit{time-inferred diffusion} \cite{Goel.2012}. This method leverages X's follower graph as well as the timestamps of the retweets to reconstruct the true retweet path. By considering the reverse chronological order of follower-followee information along with users' join dates, we infer the followership network. Follower-followee information is inferred at the time of the retweet.

The resulting dataset contains $N = \num[group-separator={,},group-minimum-digits=1]{25219}$ cascades, out of which $991$ are classified as hateful and $\num[group-separator={,},group-minimum-digits=1]{24228}$ as normal. Overall, the cascades comprise $\num[group-separator={,},group-minimum-digits=1]{91165}$ tweets (root tweets and retweets), out of which $\num[group-separator={,},group-minimum-digits=1]{22313}$ are in cascades with hateful content and $\num[group-separator={,},group-minimum-digits=1]{68852}$ are in cascades with normal content.\footnote{The data is in the supplements and  available via \url{https://anonymous.4open.science/r/hatespeech_virality-2218/}}

\subsection{Computed Variables}
\label{sec:computed_variables}


\textbf{Explanatory variables (EVs):} We compute a comprehensive set of variables at the cascade level, which will later serve as EVs in our regression model. The key EV in our analysis is given by \emph{Hateful}, which is a binary variable denoting whether the cascade resulted from hateful content (=1 if the content is hateful, and =0 otherwise). We group all remaining variables into author and content variables as follows.

\emph{Author variables: } To characterize \textquote{who} has authored the root tweets, we use the Twitter API to retrieve the number of followers and followees for each author. These variables have been shown to be important determinants for why tweets go viral \cite{Zaman.2014}. Analogous to prior work \cite{Stieglitz.2013, Zaman.2014}, we also compute variables that capture an author's activity on X, namely the tweet volume (i.e., the number of tweets divided by the account age) and the account age. Furthermore, users of public interest on X are assigned a verified status\footnote{https://help.twitter.com/en/managing-your-account/about-twitter-verified-accounts}, for which we encode a corresponding binary variable (=1 if verified, and =0 otherwise).

\emph{Content variables: } We compute variables to capture \textquote{what} is included in the tweets. Specifically, we encode a binary variable indicating whether the tweet includes a media item (=1 if it has a media item, and =0 otherwise). The media item can refer to an attached image, video, poll and/or URL. Tweets can further mention other users using \textquote{@username,} which we again encode as a binary variable (=1 if the tweet has a mention, and =0 otherwise). We also encode a binary variable whether the tweet contains hashtags (=1 if the tweet has a hashtag, and =0 otherwise). Finally, we compute the length of the raw tweet (in characters), excluding URLs, mentions, and hashtags.

\noindent\textbf{Dependent variables (DVs):} Recall that we aim to explain differences in the cascade structure of hateful vs. normal content. In order to quantify the structural differences in cascades, we compute several DVs. For this, let a cascade $i = 1, \ldots, N$ be given by a tree structure $T_{i} = (r_{i}, t_{i0}, R_{i})$ with root tweet $r_{i}$, a timestamp $t_{i0}$ of the root tweet, and a set of retweets $R_{i} = \{ (p_{ik}, l_{ik}, t_{ik}) \}_{k}$, where each retweet is a 3-tuple comprising a parent $p_{ik}$, a level of depth $l_{ik}$, and a timestamp $t_{ik}$. 

We then compute the following DVs:
\begin{itemize}
\item \emph{Cascade size $y_{i}^{\mathrm{CS}}$: } The overall number of tweets and retweets in the cascade, that is, $\abs{R_{i}} + 1$.
\item \emph{Cascade lifetime $y_{i}^{\mathrm{CL}}$:} The overall size of the time frame during which the tweet travels through the network, defined as $\max{\{ t_{ik} \}_k} - t_{i0}$.
\item \emph{Structural virality $y_{i}^{\mathrm{SV}}$: } The so-called structural virality \cite{Goel.2016} is a measure for the trade-off between cascade depth and breadth. Formally, it is defined as the average \textquote{distance} between all pairs of retweeters, that is, 
\begin{equation} \label{structvir1}
v(T_{i}) = \frac{1}{n\,(n-1)} \sum\limits_{j_1=1}^{n}\sum\limits_{j_2=1}^{n} d_{ij_1,ij_2}
\end{equation}
for a cascade $T_{i}$ with size $n$ and where $d_{ij_1, ij_2}$ refers to the distance of the shortest path between retweets ${ij}_1$ and ${ij}_2$ (similar to the Wiener index) \cite{Goel.2016}. We refer to \Cref{sec:computation_sv} for a detailed description of how we calculate the structural virality.
\end{itemize}

\renewcommand{\arraystretch}{0.8}
\begin{table}[h!] \footnotesize
\centering
\scriptsize
\sisetup{round-mode=places,round-precision=2,detect-weight=true,detect-inline-weight=math}
\begin{tabular}{lS[table-format=3.4]S[table-format=2.4]S[table-format=4.1]}
\toprule
  \textbf{Variable} & \textbf{Mean} & \textbf{Median} & \textbf{SD} \\ 
  \midrule
  \multicolumn{2}{l}{\emph{Dependent variables (DVs):}} \\
  \quad Cascade size & 3.507 & 1.000 & 58.447 \\ 
  \quad Cascade lifetime (in hours) & 58.167 & 0.000 & 1083.517 \\ 
  \quad Structural virality & 0.244 & 0.000 & 0.685 \\ 
  \addlinespace
  \multicolumn{2}{l}{\emph{Explanatory variables (EVs):}} \\
  \quad Hateful & 0.040 & 0.000 & 0.196 \\
  \quad Followers (in 1000s) & 37.689 & 0.711 & 811.427 \\ 
  \quad Followees (in 1000s) & 2.281 & 0.497 & 12.672 \\ 
  \quad Tweet volume & 27.811 & 4.850 & 83.360 \\ 
  \quad Account age (in 1000 days) & 3.552 & 3.631 & 0.981 \\ 
  \quad Verified & 0.054 & 0.000 & 0.225 \\ 
  \quad Media item & 0.736 & 1.000 & 0.441 \\ 
  \quad Mention & 0.220 & 0.000 & 0.414 \\ 
  \quad Hashtag & 0.360 & 0.000 & 0.480 \\
  \quad Tweet length & 93.994 & 92.000 & 25.380 \\ 
  \bottomrule
\end{tabular}
\caption{Summary statistics.}~\label{tbl:descriptives}
\end{table}

\subsection{Summary Statistics}\label{sec:descriptives}


\Cref{tbl:descriptives} reports summary statistics of the EVS and DVs. \textbf{DVs:} On average, a cascade has \num{2.51} retweets, lives for \num{58.17} hours, and has a structural virality of \num{0.24}. The structural properties of the cascades exhibit heavy right-skewness, as evidenced by smaller medians than means and relatively large standard deviations. \textbf{EVs:} Authors in our dataset have on average \num{37.69} thousand followers, \num{2.28} thousand followees, a tweet volume of \num{27.81}, and an account age of \num{3.55} thousand days. Out of all cascades, \num{5.4}\% originate from authors with a verified status. Several author variables are right-skewed, such as followers, followees, and tweet volume.

\begin{figure*}
	\centering
 \addtolength{\tabcolsep}{-8pt} 
	\subfloat{\label{fig:ccdfs}
    \begin{tabular}[b]{@{}ccc@{}}
        \includegraphics[width=.34\linewidth]{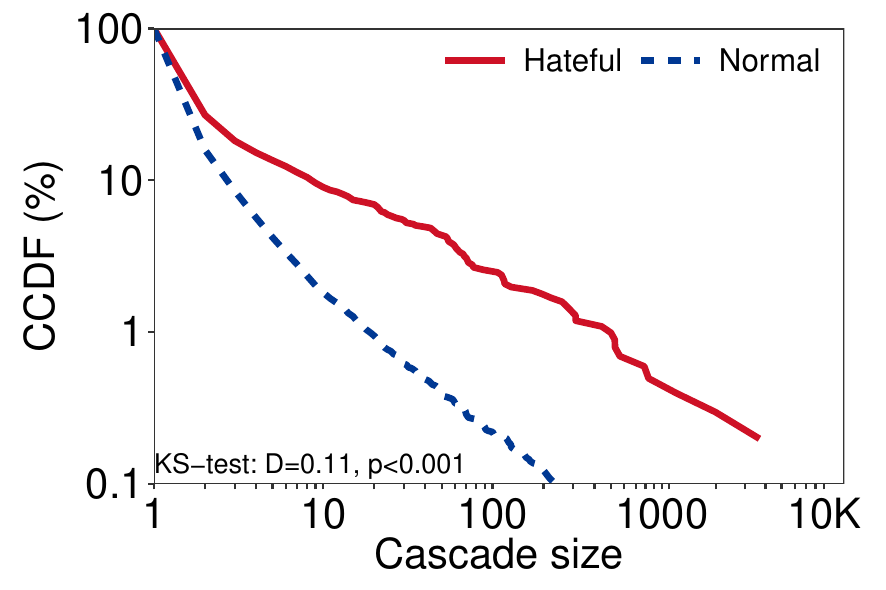}&
        \includegraphics[width=.34\linewidth]{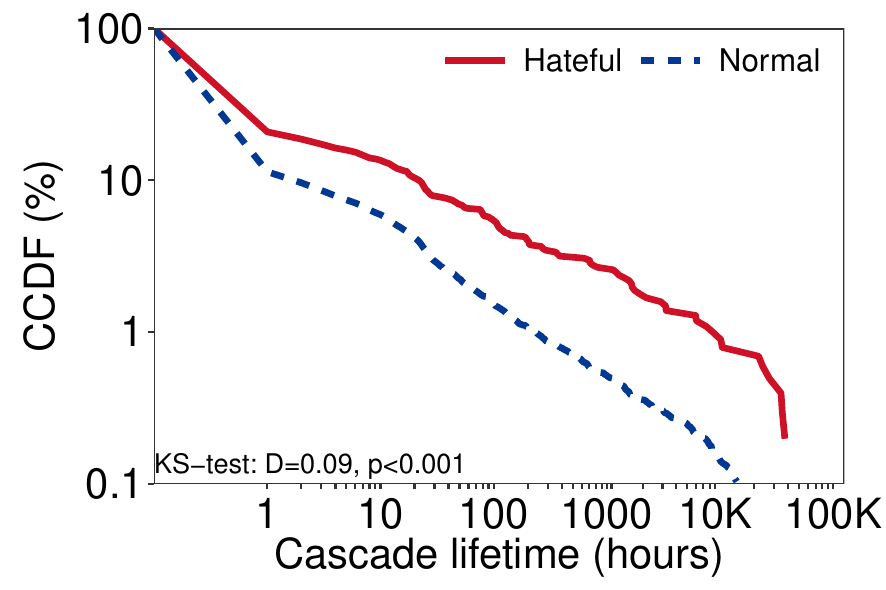}&
        \includegraphics[width=.34\linewidth]{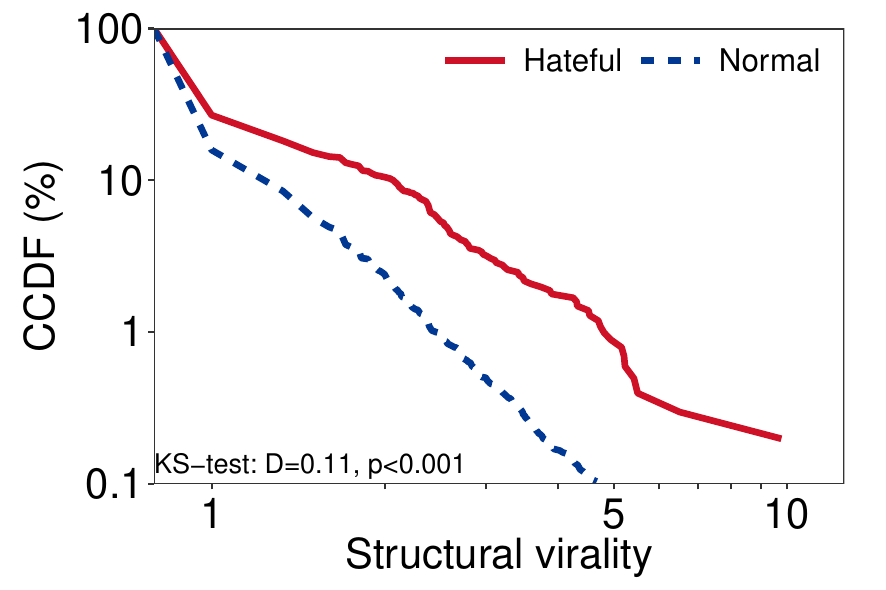}\\
    \end{tabular}%
    } \hfill
	\caption{Complementary cumulative distribution functions (CCDFs) for cascade size, cascade lifetime, and structural virality of hateful vs. normal content on social media.}
\label{fig:ccdf_dv_fs}
\end{figure*}

\Cref{fig:ccdf_dv_fs} compares the structural properties of cascades resulting from hateful vs. normal content using complementary cumulative distribution functions (CCDFs). We find that hateful content is characterized by cascades of larger size, longer lifetime, and higher structural virality. For instance, the average cascade for hateful content includes \num{22.10} retweets, has a lifetime of \num{310.16} hours, and is of \num{0.52} structural virality. For normal content, the average cascade has \num{2.73} retweets, has a lifetime of \num{47.64} hours, and is of \num{0.23} structural virality. To assess whether the differences in the distributions are statistically significant, we apply Kolmogorov-Smirnov (KS) tests.
We find that cascades with hateful vs. normal tweets show significant structural differences (all $p$-values $< 0.001$).


\subsection{Ethical Considerations}

The dataset used in this study is public, open access, and de-identified (i.e., the names of all user accounts were removed). The collection of the dataset was conducted following legal standards, as well as standards for ethical research \cite{Rivers.2014}. By accepting X’s privacy policy, non-anonymous users agree that their content and profile information is ``available for viewing by the general public''\footnote{https://twitter.com/en/privacy}.

We respect the privacy and agency of all people potentially impacted by this work and take specific steps to protect their privacy. We do not de-identify users that have chosen to remain anonymous. Further, we analyze data only in aggregate form. 

\section{Methods}

\subsection{Problem Statement}


We aim to identify determinants that explain differences in the structural properties of hateful vs. normal cascades. For this, we perform a regression analysis, which allows us to estimate and interpret \emph{marginal effects} as follows: all else being equal, how much does a specific structural property change on average if the hateful tweet features certain author or content characteristics. For example, how much larger a cascade becomes when hateful content is posted by a verified user instead of a non-verified user. To estimate such marginal effects, we build upon explanatory modeling to ensure that the model coefficients are interpretable \cite{Breiman.2001}.  Note that estimates of a regression analysis with observational data demonstrate associations and not causal paths. Thus, we later address the possibility of unobserved confounding by conducting a causal sensitivity analysis (see Sec.~\ref{sec:sensitivity}).

\subsection{Model Specification}


We follow earlier research on modeling the spread of social media content \cite{Prollochs.2021} and use generalized linear models (GLMs) to explain a specific structural property of cascades based on author and content variables. Let $y_{i}$ denote the DV in the regression; i.e., it reflects a specific structural property of cascade $i$. Analogous to earlier research \cite{Goel.2016, Prollochs.2021}, we focus on three DVs: cascade size ($y_{i}^{\mathrm{CS}}$), cascade lifetime ($y_{i}^{\mathrm{CL}}$), and cascade structural virality ($y_{i}^{\mathrm{SV}}$) .


In our GLM, the variable $\phi_i$ indicates if the content propagated in cascade $i$ is hateful ($\phi_i = 1$) or normal  ($\phi_i = 0$). Further, we include variables associated with author and content characteristics in $x_{i}$. For the list of variables included in $x_{i}$, we refer to \Cref{tbl:descriptives}.


\noindent\textbf{Regression model:}\footnote{We tested for multicollinearity using variance inflation factors. The factors for all EVs are below the critical value of five \cite{Akinwande.2015}. We also tested a model specification in which we replaced the linear relationships for the EVs with their quadratic terms to control for non-linear relationships. Here, we found that our results are robust and provide consistent support for our findings.} We estimate the following generalized linear models
\begin{align}\label{eq:regression}
    g(y_{i}) = & \, \beta_0 + \beta_1 \, \phi_i + \beta_2^T \, x_{i} + \beta_3^T \, (\phi_i \odot x_{i})
\end{align}
with intercept $\beta_0$, coefficients $\beta_1, \ldots, \beta_3$ (out of which $\beta_2$ and $\beta_3$ are vectors), link function $g(\cdot)$, and an element-wise multiplication $\odot$. Here, $\phi_i \odot x_{i}$ is a two-way interaction term. Depending on the DV (\ie, $y_{i}^{\mathrm{CS}}$, $y_{i}^{\mathrm{CL}}$, or  $y_{i}^{\mathrm{SV}}$), we choose a different underlying error distribution, link function $g(\cdot)$, and estimator. Details on the estimators for the different DVs are given in the next section.

\noindent\textbf{Model coefficients: } The estimation results for the parameters $\beta_0, \ldots, \beta_3$ characterize the spread of hateful vs. normal content as follows:
\begin{itemize}
\item $\beta_0$ is the intercept. It captures the baseline spreading dynamics. 
\item $\beta_1$ assesses the overall contribution of hateful content to spreading dynamics (all else being equal). Hence, this parameter quantifies to what extent hateful content spreads more widely (for $y_{i}^{\mathrm{CS}}$), lasts longer (for $y_{i}^{\mathrm{CL}}$), and has higher structural virality (for $y_{i}^{\mathrm{SV}}$) as compared to normal content.
\item $\beta_2$ measures how author and content variables are associated with the cascade structure. The values in $\beta_2$ do not distinguish hateful vs. normal content, and, hence, are also named \textquote{direct effects} in regression modeling.
\item $\beta_3$ estimates differences in the virality between hateful and normal cascade. It thus returns the relative differences in how variables associated with author and content characteristics are received in relation to hateful vs. normal content. This is captured by the two-way interaction between $x_{i}$ and $\phi_i$. Hence, positive values in $\beta_3$ indicate that, \eg, an increase in the number of followers is associated with a larger DV for hateful than for normal content. As we control for other variables, these estimates are \textquote{ceteris paribus.} All else being equal, they measure how much larger/smaller the effect of an author or content characteristic on size, lifetime, and structural virality is, if the tweet is hateful. Estimation results for this vector of coefficients are of particular interest, as they capture the determinants for differences in the spreading of hateful vs. normal content.
\end{itemize}

\subsection{Estimation Details}

\noindent\textbf{Estimator: } We fit generalized linear models as they allow for more flexibility regarding different error distribution compared to standard ordinary least squares. More specifically, the error is allowed to follow a distribution of the exponential family (and not just a normal distribution). In addition, the variance of the error is allowed to depend on the EVs, and, as a result, the homoscedasticity assumption of linear models is relaxed \cite{Nelder.1972}. This is beneficial for our setting as the summary statistics have shown that several of our DVs are skewed.

Each DV, i.e., $y_{i}^{\mathrm{CS}}$, $y_{i}^{\mathrm{CL}}$, and $y_{i}^{\mathrm{SV}}$, differs in its underlying distribution and therefore requires a different estimator. (1)~Cascade size ($y_{i}^{\mathrm{CS}}$) is represented by count data and the variance is larger than the mean (see \Cref{tbl:descriptives}). To adjust for overdispersion, we opt for a negative binomial regression with a log-transformation as the link-function $g(\cdot)$. (2)~Cascade lifetime ($y_{i}^{\mathrm{CL}}$) is summed to be log-normally distributed, as suggested by prior literature \cite{Zaman.2014}. Thus, we apply ordinary least squares~(OLS) with a log transformation. (3)~Structural virality ($y_{i}^{\mathrm{SV}}$) is a right skewed and non-negative continuous variable. We therefore use a gamma regression with a log link.

\noindent\textbf{Implementation:} We use R~4.2.1 for our implementation of generalized linear models. For the negative binomial regression, we use the {MASS package}.

\noindent\textbf{Interpretation: } We report 95\,\% confidence intervals for our coefficient estimates. We also report $p$-values and the corresponding significance levels. We apply $z$-standardization to all continuous EVs to facilitate interpretations. Regression coefficients $\beta_2$, $\beta_3$, and $\beta_4$ thus measure the relationship with the DV in standard deviations. Due to the log-links, the regression coefficients should be interpreted as follows: A one unit shift in the EV of interest is associated with a ${(\rm e}^{\beta_j}-1) \times 100\%$ increase in the DV.

\section{Results}

We now report the estimation results for our regression models from Eq.~(\ref{eq:regression}) and thereby identify determinants explaining differences in structural properties of hateful vs. normal cascades. In the following, we focus our discussion on coefficients $\beta_1$ and $\beta_3$ (i.e., the dummy for hateful content and the interaction terms). The reason is that these variables explain differences in the spread between hateful vs. normal content. The direct effects ($\beta_2$) have been discussed in prior works \cite{Stieglitz.2013, Zaman.2014}, and, to avoid repetition, we relegate to these works for a theoretical explanation. 

\begin{figure}[h!]
\captionsetup{position=top}
\centering
{\includegraphics[width=.5\linewidth]{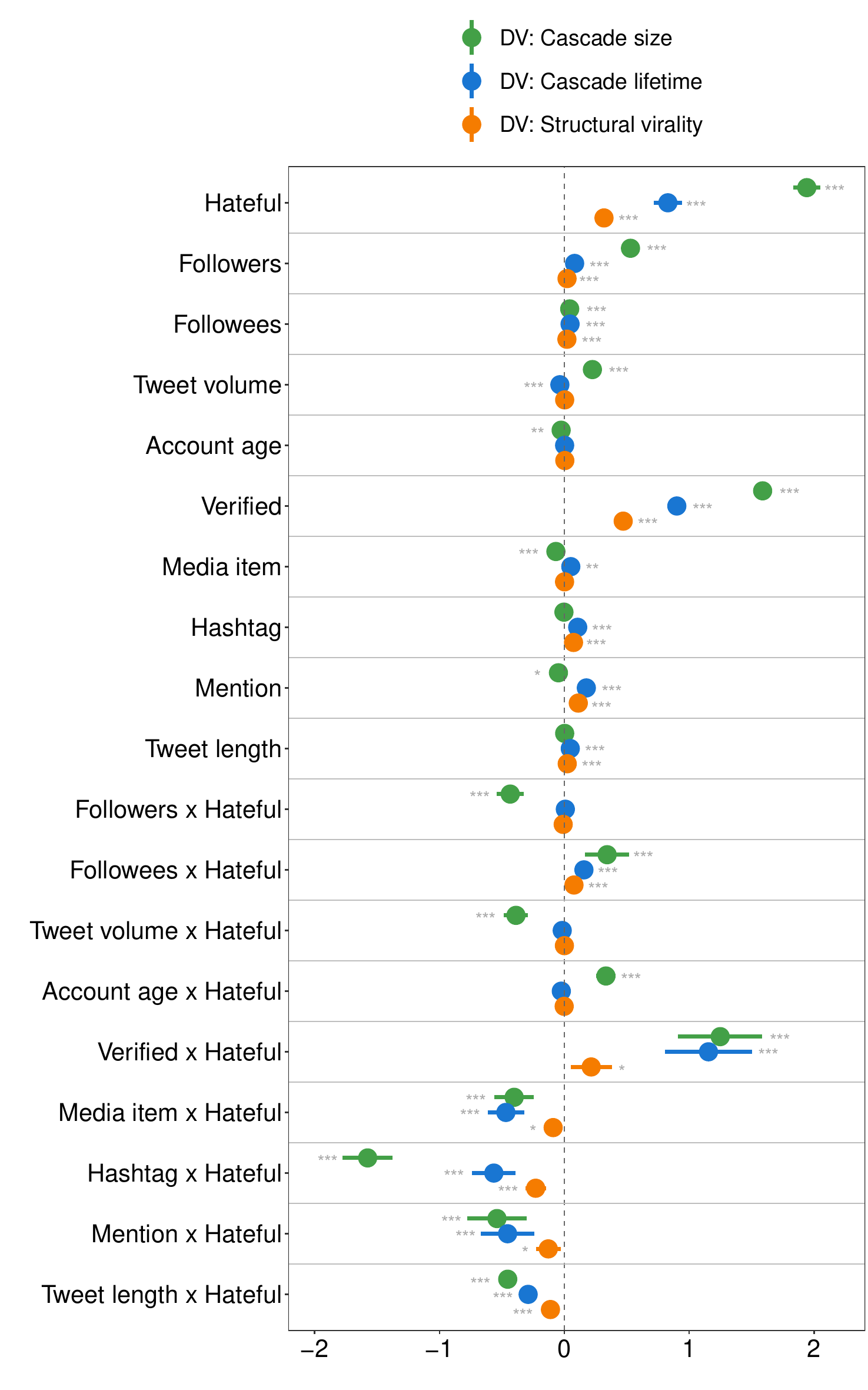}\label{fig:regression_coefs}}
\caption{Parameter estimates (standardized) and \SI{95}{\percent} confidence intervals for (a)~cascade size, (b)~cascade lifetime, and (c)~structural virality. The direct effects (top) explain the spread of all tweets (regardless of whether these embed hate or not), while the interactions (bottom) explain differences between the spread of hateful. vs. normal content. Significance levels: ***$\bm{p<0.001}$, **$\bm{p<0.01}$, and *$\bm{p<0.05}$.}
\label{fig:regression_coefficients}
\end{figure}

\vspace{0.2cm}
\noindent\textbf{Coefficients: } The coefficient plot in \Cref{fig:regression_coefficients} shows that cascades with hateful content grow larger in size, live longer, and are of larger structural virality. This is seen by the significant and positive estimates for coefficient $\beta_1$ (i.e., the dummy for hateful content) in the different regression models explaining cascade size (coef.: \num{1.941}, $p$-value $<0.001$), cascade lifetime (\num{0.829}, $p$-value $<0.001$), and structural virality (\num{0.318}, $p$-value $<0.001$). Thus, all else being equal, cascades with hateful content are \num{8.0} times larger, live \num{2.3} times longer, and have \num{1.4} times larger structural virality as compared to normal content.


As shown by the coefficient estimates, several author variables explain differences in the spreading dynamics. Out of all author variables, the largest coefficient is observed for the verified status: the verified status has a positive and significant association with cascade size (coef.: \num{1.588}; $p$-value $<0.001$), cascade lifetime (\num{0.901}; $p$-value $<0.001$), and structural virality (\num{0.471}; $p$-value $<0.001$). Moreover, the verified status explains differences in the spreading dynamics of hateful vs. normal content: the coefficient for the interaction term ($\textit{Verified} \times \textit{Hateful}$) is significant and positive for all three DVs, namely cascade size (\num{1.248}, $p$-value $<0.001$), cascade lifetime (\num{1.154}, $p$-value $<0.001$), and structural virality (\num{0.215}, $p$-value $<0.05$). Compared to normal content, hateful content stemming from a verified user is linked with a \num{3.5} times larger cascade size, a \num{3.2} times larger cascade lifetime, and a \num{1.2} times larger structural virality. Hence, the verified status appears to drive the spread of hateful content disproportionally more than normal content.

We find that changes in the number of followers and the tweet volume are associated with greater changes in size, in case of normal as opposed to hateful content: the coefficients for the respective interaction terms with the hate dummy are negative. For hateful content (compared to normal content), a one standard deviation larger number of followers is associated with a \SI[retain-explicit-plus]{35.14}{\percent} smaller cascade size ($p$-value $<0.001$). For the tweet volume, the association amounts to a \SI[retain-explicit-plus]{32.07}{\percent} smaller cascade size ($p$-value $<0.001$). The associations of both interaction terms with cascade lifetime and structural virality are not significant. In contrast, the coefficients for the interaction term $\textit{Followees} \times \textit{Hateful}$ are positive and significant for all DVs. Further, the interaction $\textit{Account age} \times \textit{Hateful}$ is positive and significant but only for cascade size. Compared to normal content, a one standard deviation larger number of followees for authors of hateful content is associated with cascades of \SI[retain-explicit-plus]{40.92}{\percent} larger size ($p$-value $<0.001$), \SI[retain-explicit-plus]{17.00}{\percent} longer lifetime ($p$-value $<0.001$), and \SI[retain-explicit-plus]{8.00}{\percent} larger structural virality ($p$-value $<0.001$). For hateful content (compared to normal content), a one standard deviation older account age is associated with a \SI[retain-explicit-plus]{39.51}{\percent} larger cascade size ($p$-value $<0.001$), while the associations with cascade lifetime and structural virality are not significant at common significance thresholds.

The findings in \Cref{fig:regression_coefficients} also identify several content variables that explain structural differences between hateful vs. normal cascades. We observe that mentions explain variations in all three structural properties (i.e., cascade size: $p$-value $<0.05$, cascade lifetime: $p$-value $<0.001$, and structural virality: $p$-value $<0.001$). More importantly, the coefficient for the interaction term $\textit{Mention} \times \textit{Hateful}$ shows that the dynamics differ for hateful vs. normal content: hateful content (as compared to normal content) with mentions of other users has a \SI[retain-explicit-plus]{41.67}{\percent} smaller cascade size, a \SI[retain-explicit-plus]{36.50}{\percent} smaller cascade lifetime, and a \SI[retain-explicit-plus]{12.03}{\percent} smaller structural virality. Thus, mentions appear to proliferate the spread of normal content, but curb the spread of hateful content.\footnote{The negative coefficient for the interaction term $\textit{Mention} \times \textit{Hateful}$ suggests that directed hate speech (i.e., hate against a specific user) is less viral than generalized hate speech (i.e., hate against a general group of individuals). As an additional check, we tested a model variant in which we replaced mentions with directed mentions, i.e., tweets which contain both mentions and second-person pronouns (e.g., you, your), as suggested by \citet{ElSherief.2018}. Our estimates are robust and yield similar results with directed mentions. For this, we counted second-person pronouns using the LIWC dictionary \cite{Pennebaker.2015}.} We find similar associations for hashtags: while the coefficient for the hashtag dummy is only significant in explaining cascade lifetime (\num{0.107}, $p$-value $<0.001$) and structural virality (\num{0.074}, $p$-value $<0.001$), the coefficient for the interaction term ($\textit{Hashtag} \times \textit{Hateful}$) is negative and significant for all three structural properties (all $p$-values $<0.001$). The use of hashtags thus appears to curb the spread of hateful content.

We also find that the length of tweets is an important determinant for the structural properties of cascades. The corresponding coefficient is positive and significant for cascade lifetime and structural virality, i.e., \num{0.05} for cascade lifetime ($p$-value $<0.001$), and \num{0.02} for structural virality ($p$-value $<0.001$), but not significant for cascade size ($p$-value $=0.76$). More importantly, the coefficient for the interaction term ($\textit{Tweet length} \times \textit{Hateful}$) is negative and significant for all three structural properties (all $p$-values $<0.001$): \num{-0.45} for cascade size, \num{-0.29} for cascade lifetime, and \num{-0.11} for structural virality. For media items, the coefficient is only significant in explaining cascade size (\num{-0.067}; $p$-value $<0.001$) and cascade lifetime (coef.: \num{0.052}; $p$-value $<0.01$). In addition, media items explain differences in the spreading of hateful vs. normal content. The coefficients for the interaction term ($\textit{Media item} \times \textit{Hateful}$) are negative and significant in explaining cascade size (\num{-0.402}; $p$-value $<0.001$), cascade lifetime (\num{-0.468}; $p$-value $<0.001$), and structural virality (\num{-0.089}; $p$-value $<0.05$). Similar to hashtags and mentions, attaching media items is associated with a less pronounced spreading for hateful content.

\begin{figure*}[h!]
\captionsetup{position=top}
\captionsetup[subfloat]{captionskip=-1pt}
\addtolength{\tabcolsep}{-6pt} 
	\subfloat[Cascade size]{\label{fig:marginal_effects_size}
\begin{tabular}[b]{@{}cccc@{}}
\includegraphics[width=.25\linewidth]{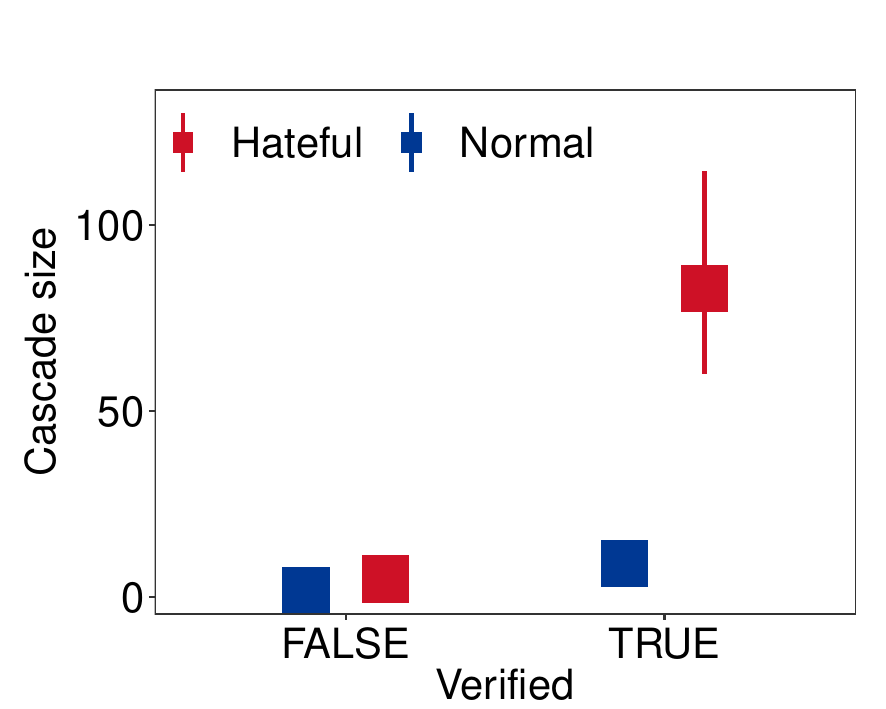}&
\includegraphics[width=.25\linewidth]{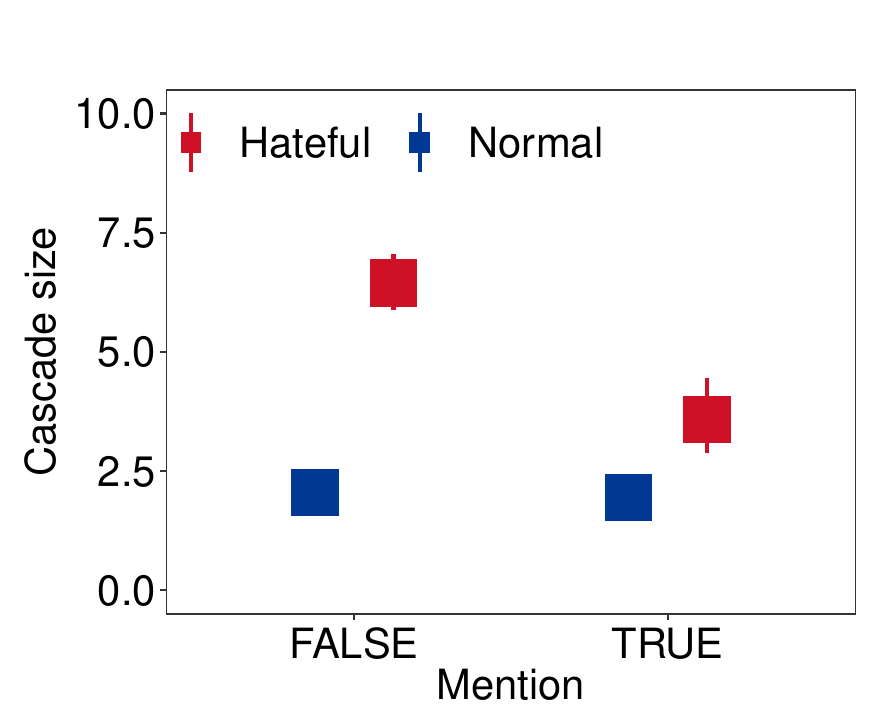}&
\includegraphics[width=.25\linewidth]{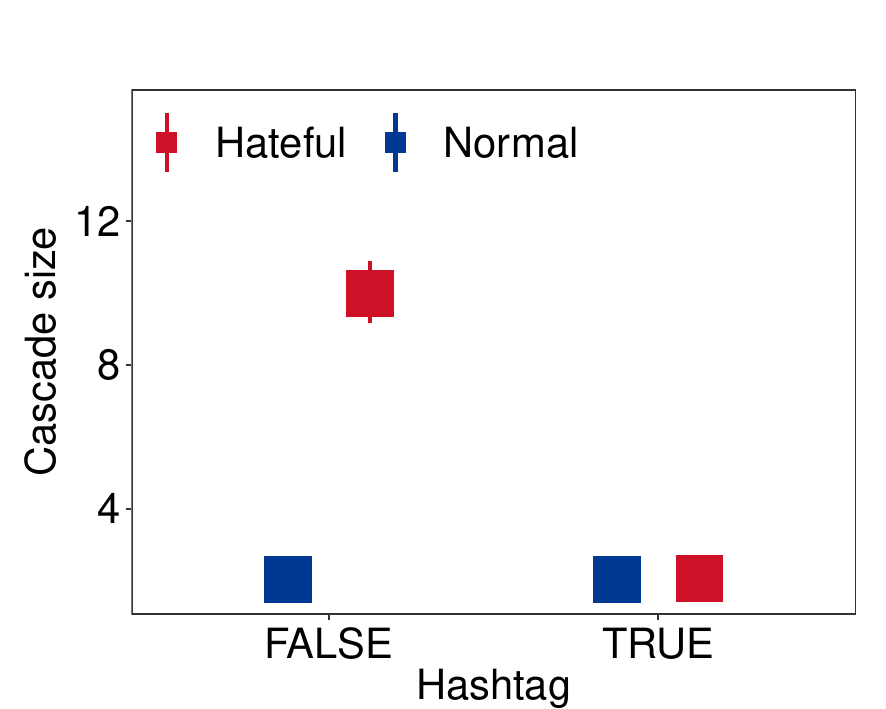}&
\includegraphics[width=.25\linewidth]{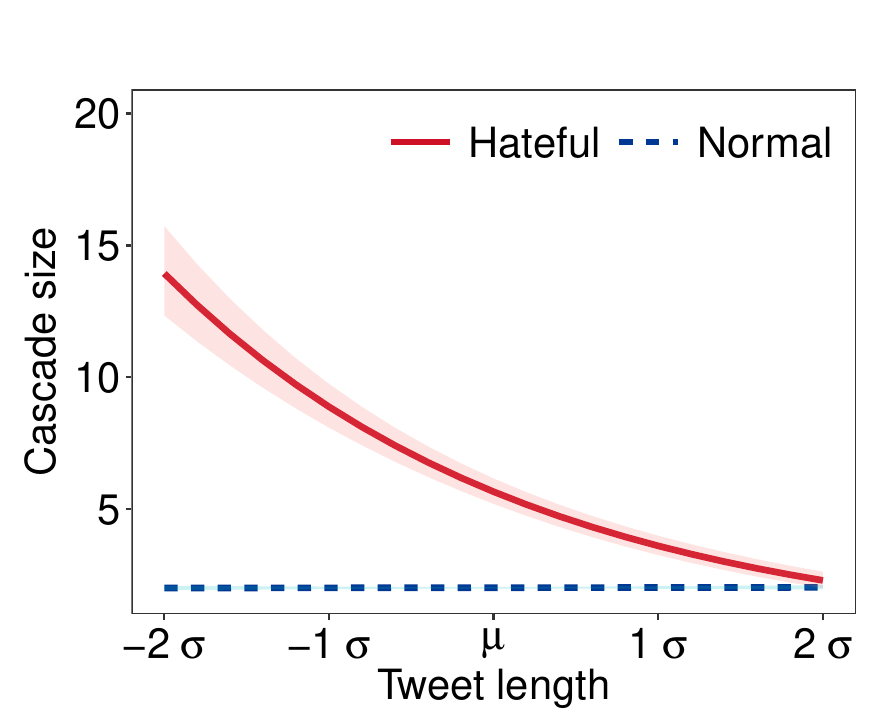}\\
\end{tabular}%
} \hfill
\subfloat[Cascade lifetime]{\label{fig:marginal_effects_liftime}
\begin{tabular}[b]{@{}cccc@{}}
\includegraphics[width=.25\linewidth]{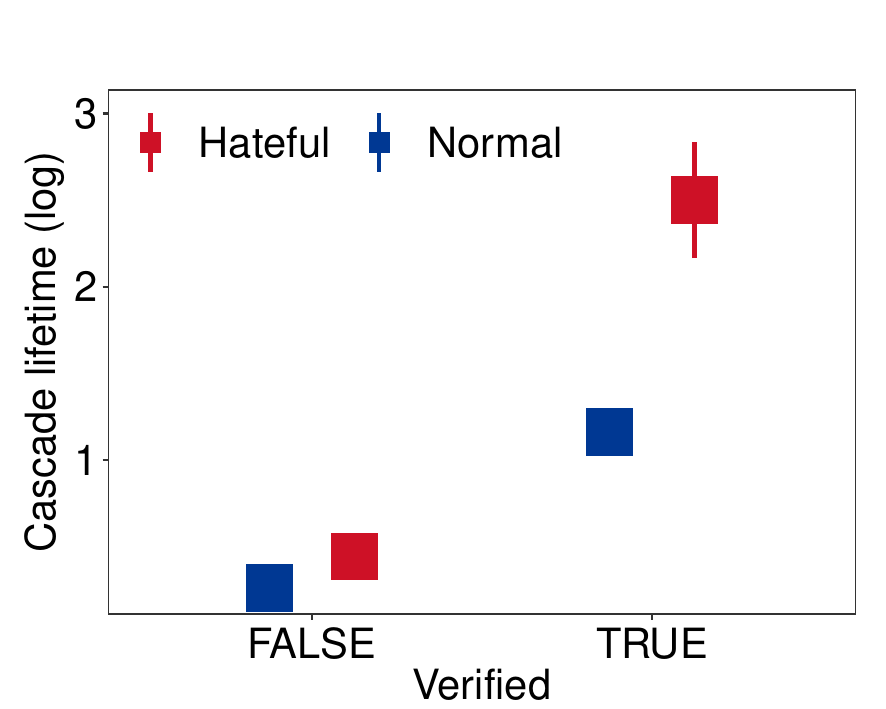}&
\includegraphics[width=.25\linewidth]{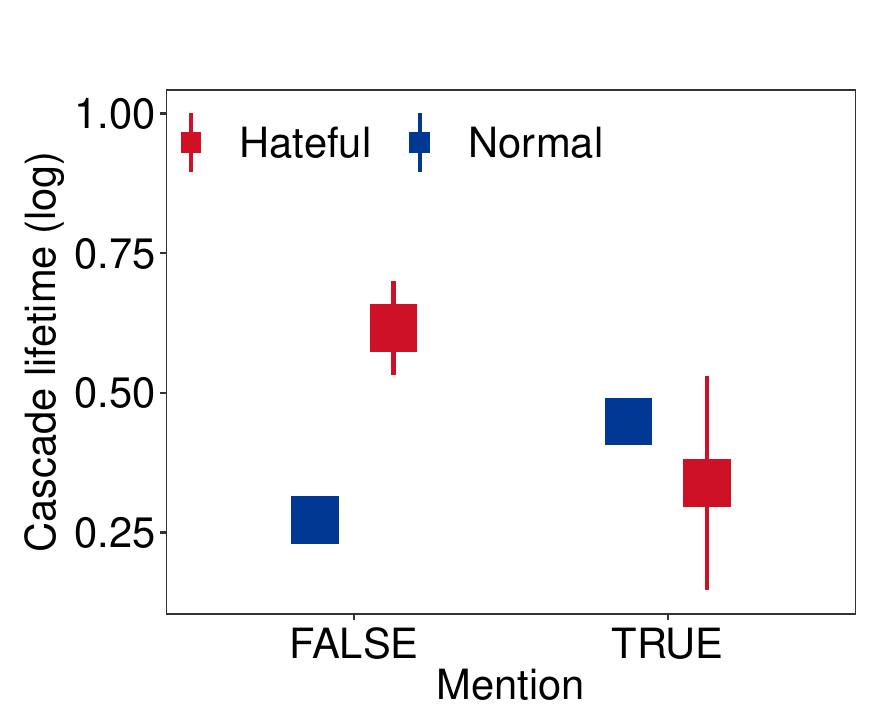}&
\includegraphics[width=.25\linewidth]{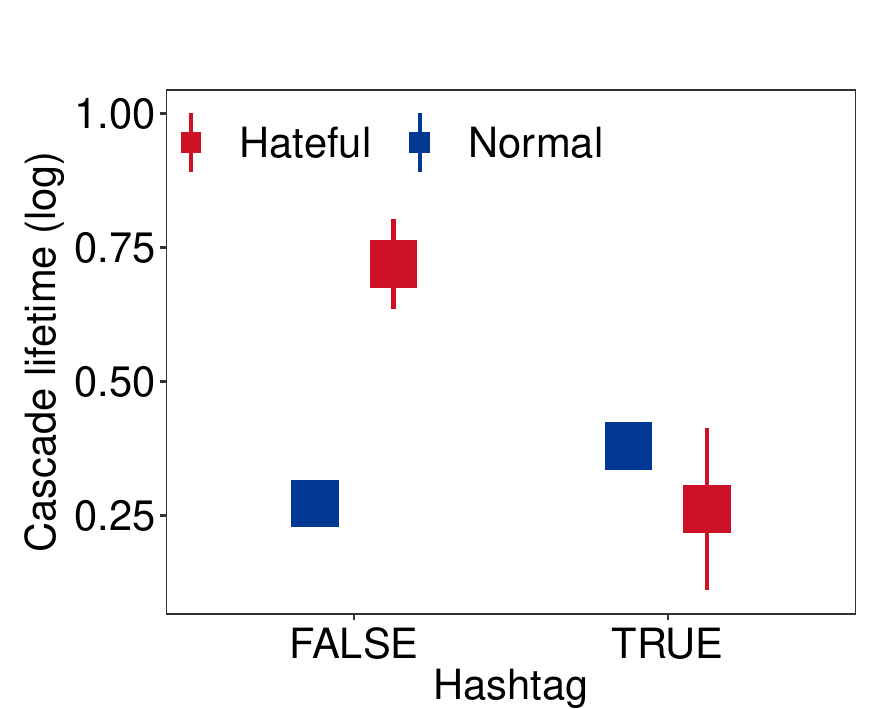}&
\includegraphics[width=.25\linewidth]{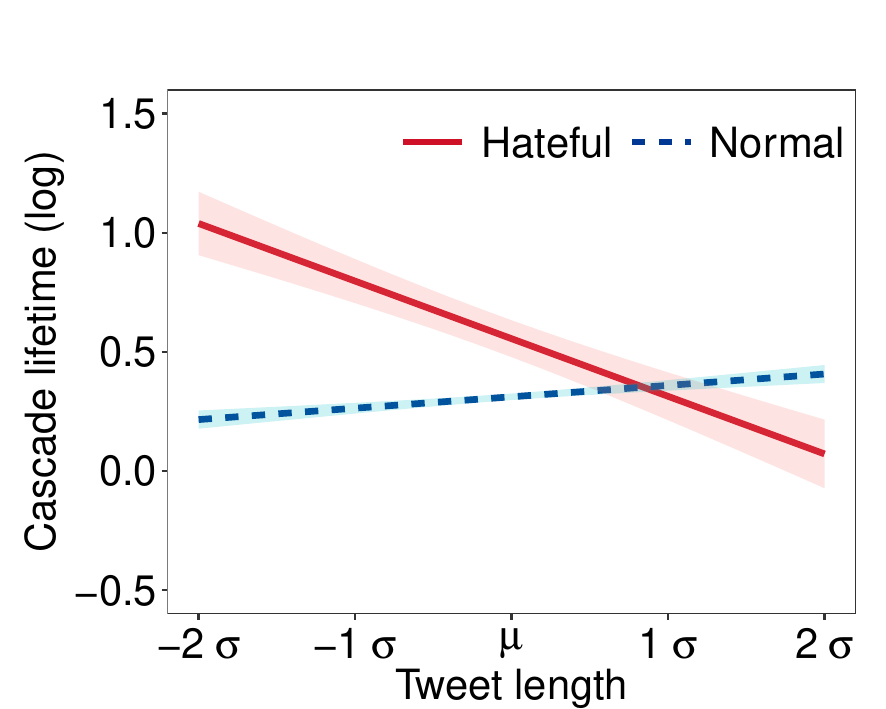} \\
\end{tabular}%
} \hfill
\subfloat[Structural virality]{\label{fig:marginal_effects_sv}
\begin{tabular}[b]{@{}cccc@{}}
\includegraphics[width=.25\linewidth]{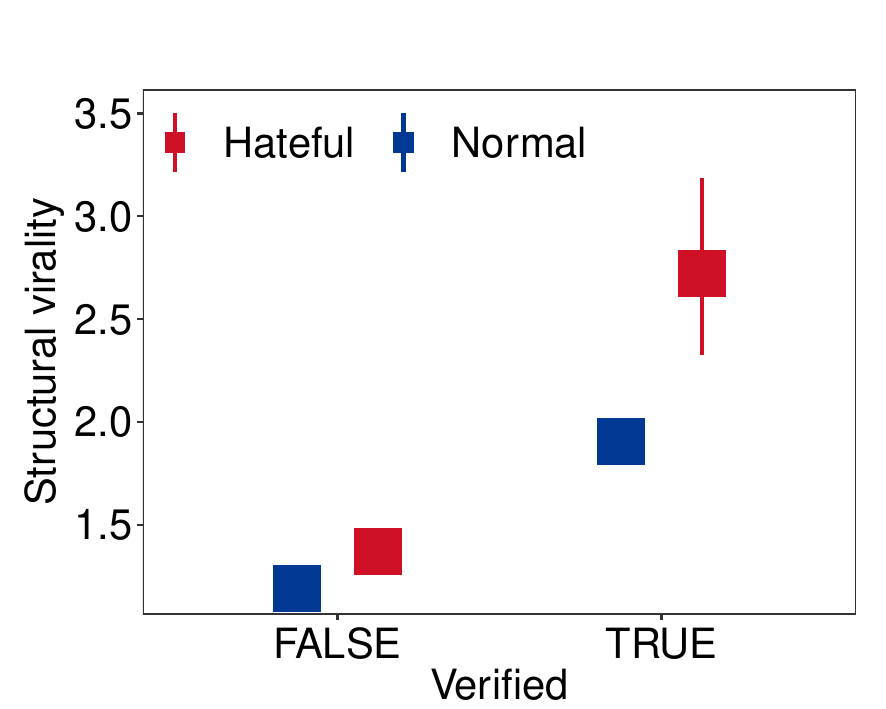}&%
\includegraphics[width=.25\linewidth]{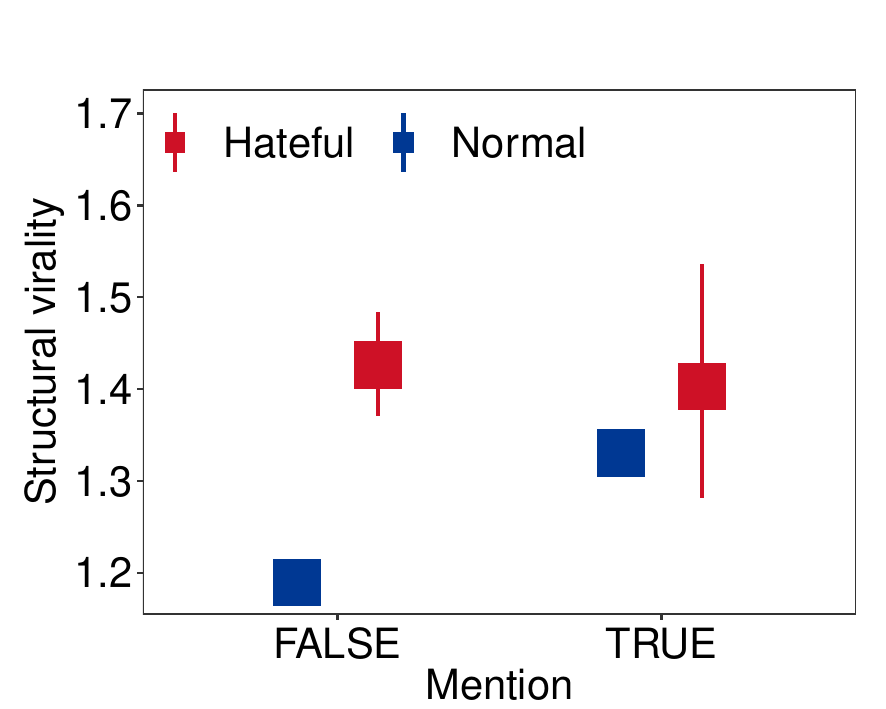}&
\includegraphics[width=.25\linewidth]{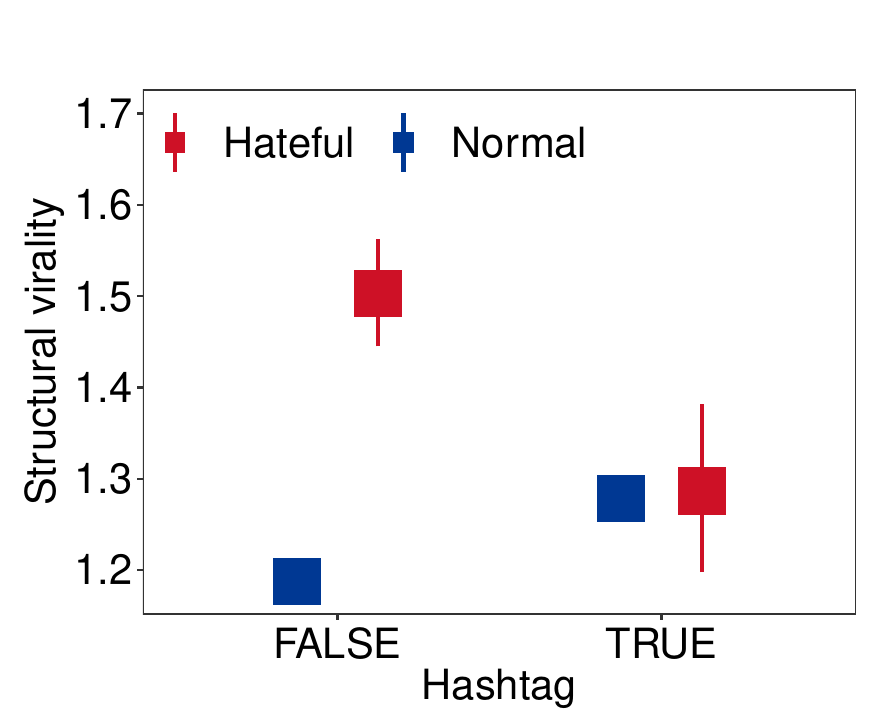}&
\includegraphics[width=.25\linewidth]{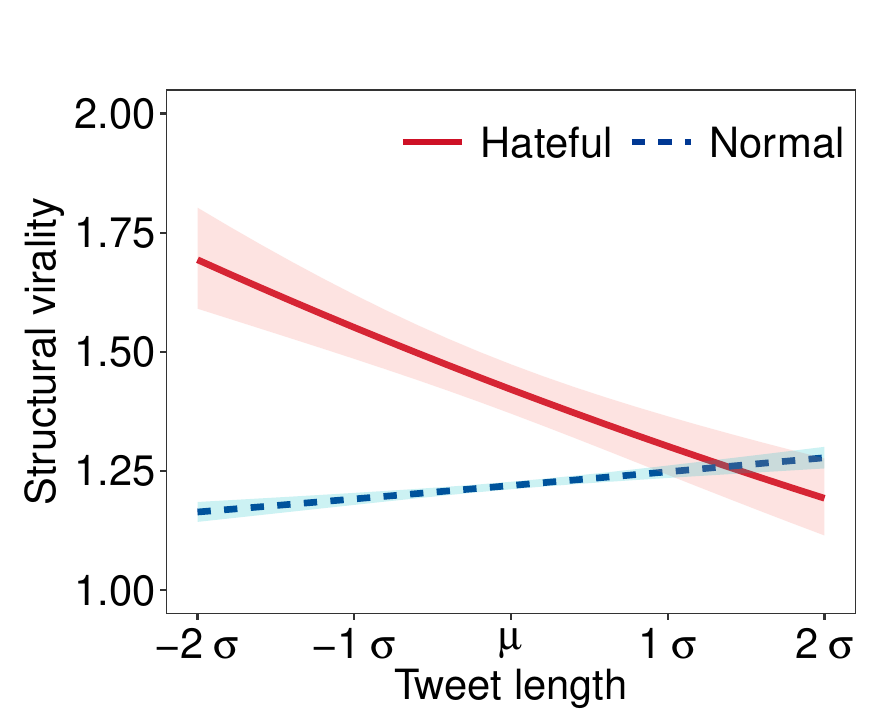} \\
\end{tabular}%
}
\addtolength{\tabcolsep}{7pt} 
\caption{Predicted marginal effects of the key determinants on (a)~cascade size, (b)~cascade lifetime, and (c)~structural virality. Shown is the mean estimate with \SI{95}{\percent} confidence intervals for hateful content (in \textcolor{BrickRed}{red}) and for normal content (in \textcolor{NavyBlue}{blue}). Continuous variables (here tweet length) were $\bm{z}$-standardized. Thus, their values on the $x$-axis represent the mean ($\bm{\mu}$) and standard deviations ($\bm{\sigma}$).
}
\label{fig:marginal_effects}
\end{figure*}

\noindent\textbf{Marginal effects: } We now analyze the marginal effects on cascade size, cascade lifetime, and structural virality. The marginal effects are computed as follows: Using the fitted regression equations, we predict the values of the DVs for varying values of an EV of interest, while holding all other EVs constant. We group the predictions into hateful vs. normal, i.e., predicted marginal effects are calculated for both hateful ($\phi_i = 1$) and normal  ($\phi_i = 0$) content, separately. This allows to further interpret differences in what makes hateful vs. normal content go viral. For reasons of space, we focus on the EVs in $x_{i}$ with the largest absolute regression coefficients: verified status, hashtags, mentions, and the tweet length (see \Cref{fig:marginal_effects}). 
In \Cref{fig:marginal_effects} (first column), we see that hateful tweets from verified authors are characterized by cascades of significantly larger size, longer lifetime, and larger structural virality (compared to hateful tweets from non-verified authors). In contrast, the estimated effects for the verified status are comparatively small for normal content.

\Cref{fig:marginal_effects} (second column) shows mixed patterns for mentions in hateful vs. normal cascades. Cascades with normal content including mentions are characterized by a slightly larger size, longer lifetime, and larger structural virality. However, all else equal, we observe the opposite for cascades from hateful content. Cascades with hateful content including mentions (compared to no mentions) are almost equal in structural virality, but smaller in size and lifetime. We find similar relationships for hashtags (\Cref{fig:marginal_effects}, third column): all else equal, normal tweets with hashtags (compared to no hashtags) are represented by cascades of slightly larger size, longer lifetime, and larger structural virality. In contrast, this pattern is not observed for hateful content: size, lifetime, and structural virality of cascades with hateful content and hashtags are similar to those of cascades with normal content. However, cascades with hateful content that include no hashtags are significantly more viral across all three structural properties. We further find that the marginal mean effects for the tweet lengths have significantly different patterns for hateful vs. normal content. All else being equal, hateful content has cascades that are of smaller size, shorter lifetime, and smaller structural virality for longer tweets. For normal content, all three structural properties are slightly larger for longer tweets.

\section{Causal Sensitivity Analysis} \label{sec:sensitivity}

We conduct a causal sensitivity analysis \citep{Cinelli.2020} to address the possibility of unobserved confounding. Thereby, we ensure that, even in the presence of unobserved confounding, our coefficient estimates cannot be explained away and thus remain robust. Specifically, we compute the robustness value (RV), which is the minimum strength of association that unobserved confounding would need to have, both with the treatment and with the outcome, to change our conclusions \citep{Cinelli.2020}. As such, the RV tells how strong a potential unobserved confounder would have to be to (a)~bring our coefficient estimates to zero ($RV_{q=1}$) and to (b) let the coefficient estimates no longer be significant at the $\alpha = 0.05$ level ($RV_{q=1, \alpha = 0.05}$).

\Cref{tbl:sensitivity} reports results of the sensitivity analysis for our generalized linear model with cascade lifetime as DV. Here, we focus on coefficient estimates for the two EVs from our hypotheses: \emph{Verified} and \emph{Mention} and their interactions with the hate dummy. For the variable \emph{Verified} (and the interaction $\emph{Verified} \times \emph{Hateful}$): (a)~an unobserved confounder would have to explain 16.2\% (and 4\%) of the residual variance for the estimate to become zero ($RV_{q = 1}$), and (b)~would have to explain 15.2\% (and 2.8\%) for the effect to be no longer significant ($RV_{q = 1, \alpha = 0.05}$). For the variable \emph{Mention} (and the interaction $\emph{Mention} \times \emph{Hateful}$): (a)~$RV_{q = 1}$ would have to be 6.2\% (and 2.6\%), and (b) $RV_{q = 1, \alpha = 0.05}$ would have to be 5\% (and 1.4\%). All of these values are larger than the strength of association of the variable \emph{Followers}, and, hence, a potential unobserved confounder must be more important than the number of followers. However, the number of followers captures the social influence of users and has thus been found to be one of the most important determinants in explaining the spread of social media both empirically and theoretically \cite{Zaman.2014}. Hence, according to social science theory, there is unlikely any confounder that can explain away our conclusions, which confirms the robustness of our findings.

\begin{table}[!h]
\centering
\scriptsize
\begin{tabular}{lrrr}
\toprule
\textbf{Treatment} & $\bm{R^2_{Y \sim D |{\bf X}}}$ & $\bm{RV_{q = 1}}$ & $\bm{RV_{q = 1, \alpha = 0.05}}$  \\ 
\midrule
\emph{Verified} & 3.1\% & 16.2\% & 15.2\% \\
\emph{Mention} & 0.4\% & 6.2\% & 5\% \\
\emph{Verified} $\times$ \emph{Hateful} & 0.2\% & 4\% & 2.8\% \\
\emph{Mention} $\times$ \emph{Hateful} & 0.1\% & 2.6\% & 1.4\% \\
\bottomrule
\multicolumn{4}{l}{$R^2_{Y \sim D |{\bf X}}$ is the partial $R^2$ of the treatment with the outcome}
\end{tabular}
\caption{Sensitivity analysis of our generalized linear model with cascade lifetime as DV.}~\label{tbl:sensitivity}
\end{table}

\section{The Role of Infectiousness as Mediator} \label{sec:infectiousness}

The different structural properties -- i.e., cascade size, cascade lifetime, and structural virality -- are unlikely to be independent. Rather, prior research has already made the observation that cascades of larger size also live longer and are more viral \cite{Juul.2021}. Hence, differences in the size of cascades can be expected to explain differences in their lifetime and structural virality. Motivated by this, we now analyze to what extent hate explains differences in cascade lifetime and structural virality \emph{beyond} differences in cascade size. Here, we follow the framework by \citet{Juul.2021} and now account for cascade size in addition to the other variables. Importantly, \citet{Juul.2021} argue that this has a meaningful interpretation: cascade size should be seen as a proxy for how online content propagates through the network and, therefore, captures the person-to-person \textquote{infectiousness} of online content, that is, its contagiousness. A larger cascade size thus indicates that more users are exposed to (\textquote{infected}) the content. In line with this, we now refer to cascade size as a proxy for the \textquote{infectiousness} (or contagiousness) of online content. Motivated by this, we now seek to understand the role of infectiousness in explaining differences in the spreading dynamics of hateful vs. normal content.  

To revise our regression model, the first question is whether infectiousness should act as a moderator or as a mediator. In statistics, a moderator variable affects the strength and direction of a relationship, while a mediator variable explains the mechanism through which two variables are related \cite{Frazier.2004}. This has also downstream implications during modeling: one accommodates a moderator simply by controlling for it analogous to any other control variable, while one accommodates a mediator through a tailored mediation analysis \cite{Shrout.2002}. Informed by prior theory, we expect the infectiousness of online content to change the process through which tweets propagate. Specifically, online spreading processes are known to be self-exciting (e.g., \cite{Crane.2008, Mishra.2016, Zhao.2015}). This refers to a mathematical property whereby the occurrence of past tweets makes sharing and thus the occurrence of future retweets more probable (thus making it more contagious). Hence, we treat infectiousness as a mediator.  

We perform a mediation analysis to understand the role of \textquote{infectiousness} as a mediator. In a mediation analysis, one seeks to understand the \emph{mechanism} that underlies a certain relationship between an EV and DV (here: between author/content variables and cascade properties). However, unlike a standard regression, we now assume a structural causal model that allows for two different causal pathways \cite{Baron.1986}: (1)~a direct relationship whereby the author/content variables directly explain variations in the cascade properties, and (2)~an indirect relationship that goes through another hypothetical variable known as mediator. In presence of a mediator, author/content variables may first promote the infectiousness of a tweet, which, in turn, then drives the lifetime and the structural virality of the cascade.  Note that a mediation analysis \emph{cannot} learn causal pathways \citep{Baron.1986}; however, it is effective for decomposing effects as is the aim in our work. Discerning the different pathways of (1) and (2)  contributes to our theoretical understanding of why hate speech goes viral. On the one hand, finding empirical support of (1) would mean that people propagate hate speech due to specific characteristics of the author or the content. On the other hand, finding empirical support of (2) would mean that people share hate speech neither because of the author nor because of the content but simply because hate speech is \textquote{infectious} (i.e., contagious). Hence, curbing the infectiousness of hate speech would be a simple way to counter its spread. 

We follow common practice for performing a mediation analysis \cite{Shrout.2002}. First, we estimate the total direct effect of hate as our EV on the DVs and on the mediator infectiousness (i.e., captured by cascade size as suggested in \cite{Juul.2021}). We control for the direct effect of all other EVs and thereby isolate the effect of hate-only. Here, we pay attention to whether the link is statistically significant. On the one hand, consistent results would corroborate the findings from our main regression. On the other hand, this is a requirement for mediation to occur. Second, we include the mediator in the regression equations to quantify the effect of infectiousness on cascade lifetime and structural virality, respectively. We expect that infectiousness explains more variance than the EV for hate. This would then provide evidence supporting a mediation. Here, one distinguishes two cases: (1)~If the regression coefficient of the EV for hate is no longer significant, the infectiousness \textquote{fully mediates} it. (2)~If the regression coefficient shrinks but stays significant, an \textquote{partial mediation} is present. Third, we use boostrapping to perform significance testing for infectiousness as mediator. For each bootstrapped sample, we calculate the causal mediation effect (CME), which is the product of the regression coefficient of the EV for hate on the mediator and the regression coefficient of the mediator on cascade lifetime (or structural virality). Using the results for each sample, we compute the average causal mediation effect (ACME), the average direct effect (ADE, i.e., the average of the coefficients of the EV for hate after controlling for infectiousness), and the 95\% confidence intervals.

\begin{figure}[h!]
\captionsetup{position=top}
\centering
\addtolength{\tabcolsep}{5pt}
\begin{tabular}[b]{@{}cccc@{}}
\subfloat[Cascade lifetime]{\includegraphics[width=0.45\linewidth]{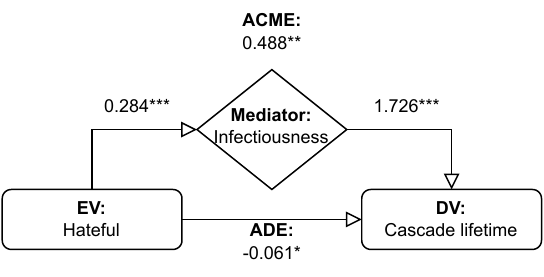}\label{fig:mediation_cl}} &
\subfloat[Structural virality]{\includegraphics[width=0.45\linewidth]{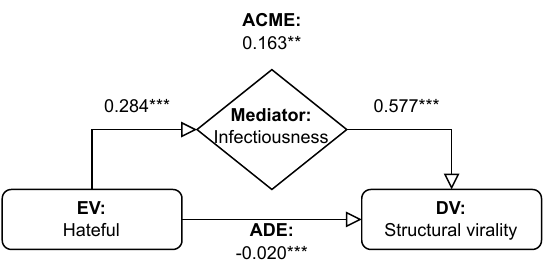}\label{fig:mediation_sv}}
\end{tabular}%
\caption{Mediation analysis captioning the average direct effect (ADE) and the average causal mediation effect (ACME) through infectiousness (mediator) of the EV (i.e., hate) on (a)~cascade lifetime and (b)~structural virality (DVs). Significance levels: ***$\bm{p<0.001}$, **$\bm{p<0.01}$, and *$\bm{p<0.05}$.}
\label{fig:mediation}
\end{figure}

The results of the mediation analysis are shown in \Cref{fig:mediation}. We first discuss the results for cascade lifetime. Evidently, the relationship between hate and cascade lifetime is partially mediated by the infectiousness. The coefficients capturing direct effects of hate on both infectiousness and cascade lifetime without infectiousness as EV are statistically significant (coef.: \num{0.284} for infectiousness, $p$-value $<0.001$; \num{0.442} for cascade lifetime, $p$-value $<0.001$). With infectiousness as EV, the coefficient capturing the direct link between hate and cascade lifetime is still statistically significant but lower in magnitude (\num{-0.061}, $p$-value $<0.05$). Thereby, the coefficient for the direct link between infectiousness and cascade lifetime is statistically significant (coef.: \num{1.726}, $p$-value $<0.001$). The results for structural virality are as follows. We again observe a partial mediation. The links hate--infectiousness and hate--structural virality without infectiousness as EV are statistically significant (coef.: \num{0.284} for infectiousness, $p$-value $<0.001$; \num{0.148} for structural virality, $p$-value $<0.001$). With infectiousness as EV, the direct link hate--structural virality is still significant but lower in magnitude (\num{0.020}, $p$-value $<0.001$), while the direct link infectiousness--structural virality is statistically significant (coef.: \num{0.577}, $p$-value $<0.001$). To this end, we perform significance testing using \num{500} bootstrapped samples. The ACMEs for cascade lifetime (\num{0.488}) and structural virality (\num{0.163}) are both statistically significant ($p$-value $<0.01$). Altogether, this confirms infectiousness as an important mediator in the underlying spreading dynamics of hate speech. 

\section{Discussion}

\textbf{Relevance:} Hate speech presents a widespread threat to the well-being of individuals \cite{Saha.2019, Muller.2021}, and it further propels the ongoing segregation and radicalization in many societies \cite{Allport.1954, Bilewicz.2020}. A particular characteristic of hate speech is that it encourages violence against vulnerable or marginalized groups, which has been repeatedly documented in hate crimes across the world \cite{Roose.2021, Mozur.2021, Taub.2021}. Here, we contribute to computational social science by advancing our understanding of what makes hate speech viral.


\noindent\textbf{Summary of findings: } We found empirical evidence that cascades with hate speech grow larger in size, live longer, and are of larger structural virality, as compared to cascades with normal content. Moreover, we identified several determinants explaining why hateful content is more viral than normal content. For example, a disproportionally larger size, lifetime, and structural virality is found for hateful content (as opposed to normal content) when the corresponding tweet comes from verified users and/or has no mentions. 
We further conducted a causal sensitivity analysis to address the possibility of unobserved confounding. Here, we found that a potential unobserved confounder must be more
important than the number of followers (\ie, one of the most important drivers of diffusion)  to explain away our conclusions. This alleviates endogeneity concerns and demonstrates the robustness of our analysis.

\noindent\textbf{Theoretical explanation:} Prior literature offers a theoretical explanation why hate speech is disproportionally viral. Hate is an emotion of particularly strong arousal and negative valence \cite{Plutchik.2001}. As such, it is known to provoke strong reactions, oftentimes encouraging violence or other defamatory behavior against others (e.g., racism). Unlike positive emotions, negative emotions are known to drive online behaviors such as interactions and sharing \cite{Robertson.2023, Stieglitz.2013, Prollochs.2021b, Prollochs.2021}. This observation has been previously stipulated in the theory on the negativity effect \cite{Baumeister.2001}. Together, this gives a behavioral explanation for why users interact with online hate speech and, thus, why online hate speech is \textquote{infectious.} 

Our finding related to the verified status can be explained by social science theory. Hate speech allows individuals to differentiate ``us'' from ``them'' and, therefore, primarily targets at out-group members \cite{Halevy.2008}. Defining in- vs. out-group members is shaped by social norms for which societal leaders play a central role \cite{Siegel.2020}. Hence, if individuals with a special social status make hate speech appear socially acceptable, it is likely that other users embrace the corresponding social norms -- and thus propagate the hateful content. Our findings for mentions and hashtags imply significant differences in the spread of directed vs. generalized hate speech on social media. Directed hate speech is directed against a specific user or entity, which differs from generalized hate that is against a general group of individuals \cite{ElSherief.2018b}. As such, generalized hate may proliferate more widely as it has the potential to appeal to a large number of users. In line with this, we find that tweets with generalized hate speech (characterized by no mentions/hashtags) are more viral than directed hate speech (characterized by the presence of mentions/hashtags). Differences between generalized vs. targeted hate have been conceptualized in prior theory \cite{ElSherief.2018b}, while we offer large-scale empirical evidence that these modulate the spread. 


\noindent\textbf{Implications:} Several regulatory and policy efforts around the world force social media platforms to limit the spread of online hate speech. Examples are, e.g., the code of conduct for social media platforms in the European Union \cite{Wigand.2020} as well as hate speech policies that have been put into effect by the platforms themselves (e.g., Facebook, YouTube). In this regard, our findings emphasize the importance of taking actions against spreaders of hate speech that have a verified status, as their high social influence amplifies the spread. Hence, tweets by verified users should be prioritized in manual assessments with the purpose of detecting online hate. Our findings also offer an explanation why automated hate speech detection is computationally challenging: viral hate speech is characterized by short tweets with no mentions and no hashtags. Hence, the absence of informative, content-based features can make it difficult for machine learning tools to detect hate speech. In contrast, our findings show that hate speech is characterized by distinctive features that manifest in unique structural properties of retweet cascades. Hence, a viable alternative to detect hate speech cascades can be to leverage cascade-based features. Similar approaches have been developed for fake news \cite[e.g.,][]{Naumzik.2022} but not for online hate speech, thereby presenting an interesting avenue for future research.


Our findings also offer recommendations for policy-makers and law enforcement authorities interested in curbing the spread of online hate speech. Our results suggest that it is particularly important to reduce the \textquote{infectiousness} of hateful content, i.e., how contagious online hate speech is. As such, policy-makers should strive to improve the online literacy in the wider public. In practice, this could be achieved by public campaigns that directly curb the infectiousness of hate speech. Examples are to inform about the negative effects for individuals targeted by hate speech and thus to raise awareness, as well as to elucidate the corresponding legal consequences for sharing hate speech. In line with this, experimental evidence has found that interventions in form of counterspeech are effective in confronting spreaders of hate speech and thereby lead to a deletion of hate speech and a subsequent reduction in hate speech creation \cite{Hangartner.2021}. Counterspeech is especially interesting for practice as such interventions are both scalable (e.g., via automated bots) and deployable by a variety of stakeholders (e.g., nongovernmental organizations such as the Center for Countering Digital Hate\footnote{https://counterhate.com/}). 

\noindent\textbf{Limitations and future research:} As with other research, ours is not free of limitations that offer opportunities for future research: (1)~We analyze the spread of hate speech on X. Thus, future research could extend our work and examine whether the findings generalize to other online social media platforms. However, comparing our descriptive findings from \Cref{sec:descriptives} with the findings on Gab in \cite{Mathew.2019} suggests that the findings are likely to be similar. Specifically, hate speech on both X and on Gab is associated with a larger cascade size, longer lifetime, and higher virality, as compared to normal content, which hints that also the mechanisms of how users share hate speech are similar for both. (2)~X has experienced several policy changes in the past year, including the option to purchase a verified status. Before that, the verified status was manually assigned to the most influential users on X. Thus the question arises as to whether the appearance of higher societal status for verified users still remains under the new policy. Future research could examine whether the new mechanism of receiving a verified status changes the dynamics of retweeting hateful vs. normal content. Specifically, future research could examine whether hateful content from verified authors is still disproportionally more likely to go viral under the new policy. (3)~The dataset from \cite{Founta.2018} comprises English language tweets only. However, there might be cultural differences (e.g., for the Global South) affecting both the spreading behavior as well as determinants for the spreading of hateful vs. normal content. As a result, taking a cross-lingual perspective presents a promising avenue for future research.

\section{Conclusion}

Social media provides new forms of connecting individuals but is also accompanied by a massive proliferation of hate speech, thus presenting an alarming consequence for marginalized and vulnerable groups, as well as society as a whole. However, little is known about the spreading dynamics of online hate speech and, in particular, what makes it viral. We found that cascades with hate speech are particularly contagious (``infectious''): they grow larger in size, live longer, and are of larger structural virality. We further identified important determinants behind the differences in the spread of hateful vs. normal content on X. To the best of our knowledge, this is the first work explaining the virality of online hate speech. Finally, our results also suggest strategies that are effective at curbing the proliferation of online hate speech.

\section*{Author Contributions}

All authors contributed to conceptualization, results interpretation, and manuscript writing. Abdurahman Maarouf contributed to data analysis. All authors approved the manuscript.

\clearpage

\bibliographystyle{ACM-Reference-Format-no-doi-abbrv}
\bibliography{literature}

\clearpage
\appendix

\vspace{1cm}
\begin{center}
\huge Supplementary Materials
\end{center}
\vspace{1cm}

\counterwithin{figure}{section}
\counterwithin{table}{section}

\section{Computation of Cascade Metrics}\label{sec:computation_sv} \noindent 

\textbf{Structural virality:} The structural virality is based on the \textit{Wiener Index}. It is defined as the average distance of all pairs of nodes in a cascade and thereby describes the trade-off between cascade depth and cascade breadth without the need to compute either of both. Formally it can be shown that the formula for structural virality can be expressed in terms of the sizes of all cascade subtrees \cite{Goel.2016}, i.e.,
\begin{equation} \label{structvir2}
v(T_{i}) = \frac{2n}{(n-1)} \left[\frac{1}{n}\sum\limits_{S\in\mathbb{S}_i}^{} |S| - \frac{1}{n^2} \sum\limits_{S\in\mathbb{S}_i}^{} |S|^2 \right]
\end{equation}
for a cascade $T_i$ with size $n$ and where $\mathbb{S}_i$ refers to the set of all subtrees in cascade $T_i$ and where $|S|$ refers to the size of the subtree. Subtrees are defined to have their origin at any node of the original cascade (but not at an end node). In our work, we make use of the formula in \Cref{structvir2} to compute the structural virality.

\begin{figure}[h!]
\captionsetup{position=top}
\centering
{\includegraphics[width=.4\linewidth]{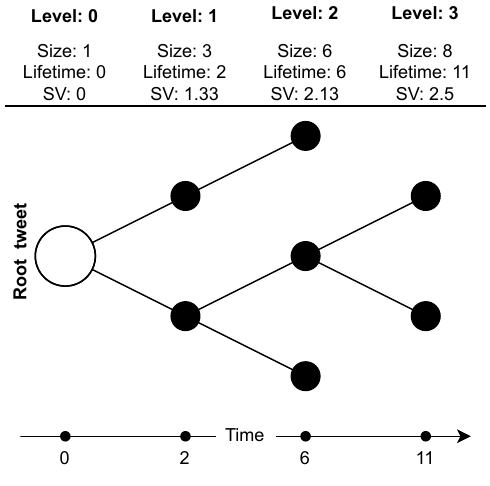}}
\caption{Exemplary cascade with computed cascade metrics at each level.}
\label{fig:example_cascade}
\end{figure}

\textbf{Cascade structure: }\Cref{fig:example_cascade} illustrates an exemplary cascade structure of content propagation on X. The root node is the original tweet (root tweet) containing either hateful or normal content. The children are retweets of the original tweet while all other nodes are retweets of retweets. We use the retweet path to calculate the cascade metrics at each level, namely, the cascade size (the number of nodes), cascade lifetime (the time difference between the root tweet and the terminal tweet), and structural virality. Note that the final cascade metrics we use for our work are those computed at the last level.

\clearpage
\section{Marginal Effects for Additional Variables} \label{sec:additional_mes} \noindent 

\begin{figure*}[hbt!]
\captionsetup{position=top}
\captionsetup[subfloat]{captionskip=-1pt}
\centering
\addtolength{\tabcolsep}{-6pt}
	\subfloat[Cascade size]{\label{fig:marginal_effects_size_2}
\begin{tabular}[b]{@{}cccc@{}}
\includegraphics[width=.25\linewidth]{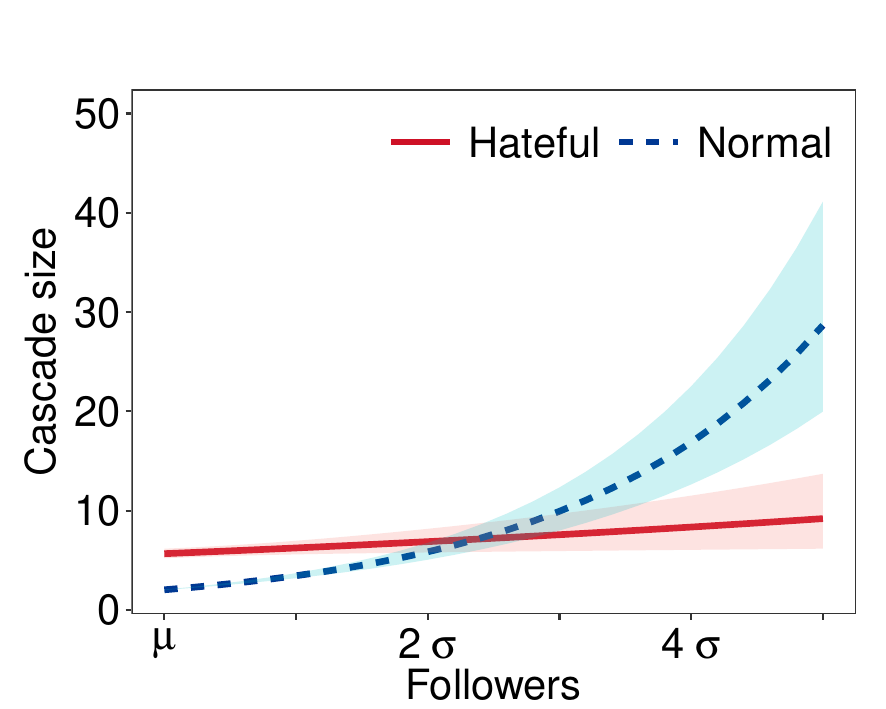}&
\includegraphics[width=.25\linewidth]{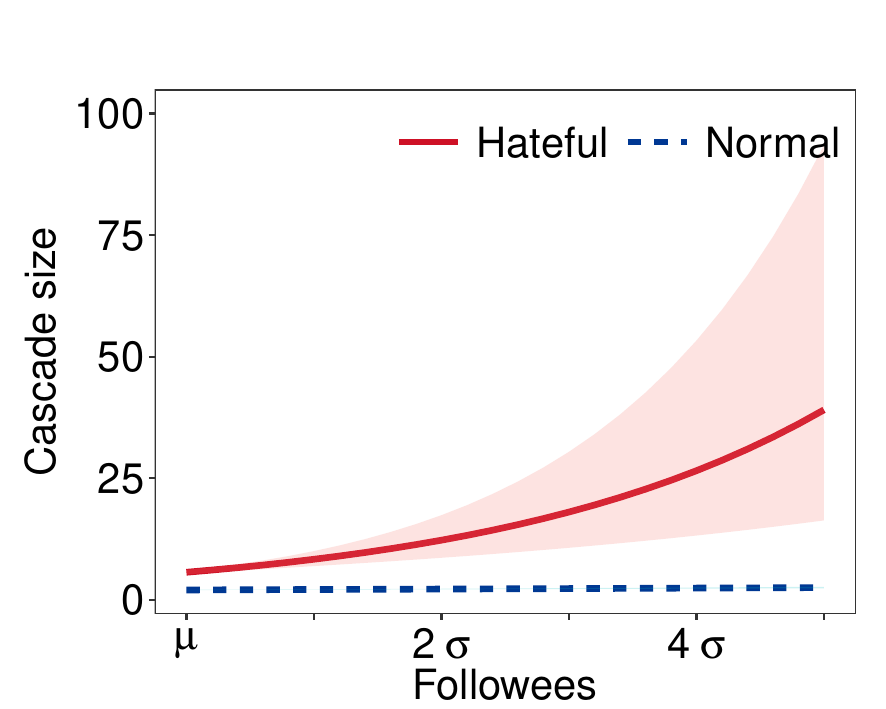}&
\includegraphics[width=.25\linewidth]{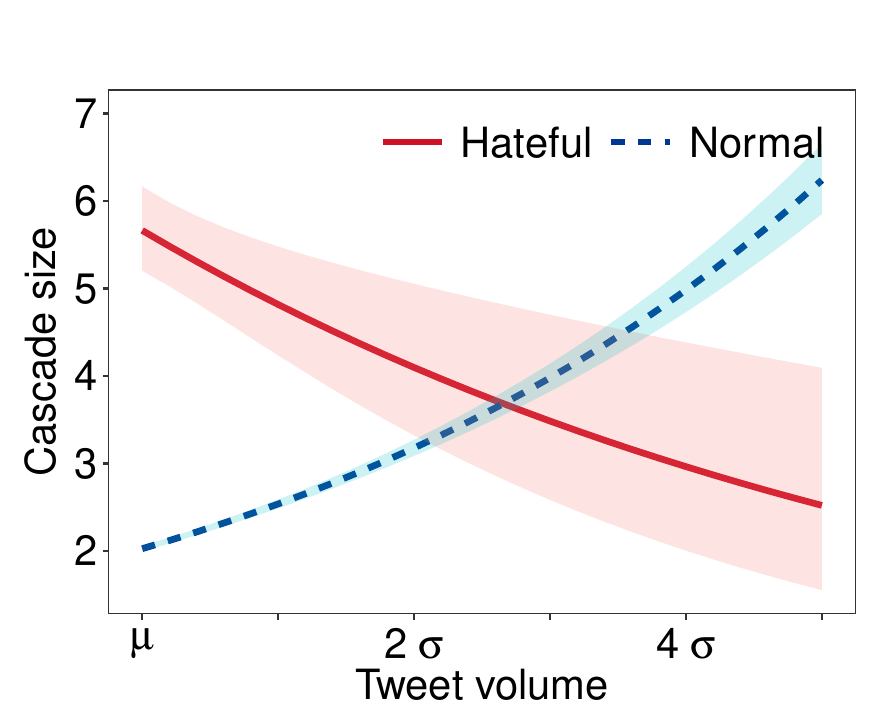}&
\includegraphics[width=.25\linewidth]{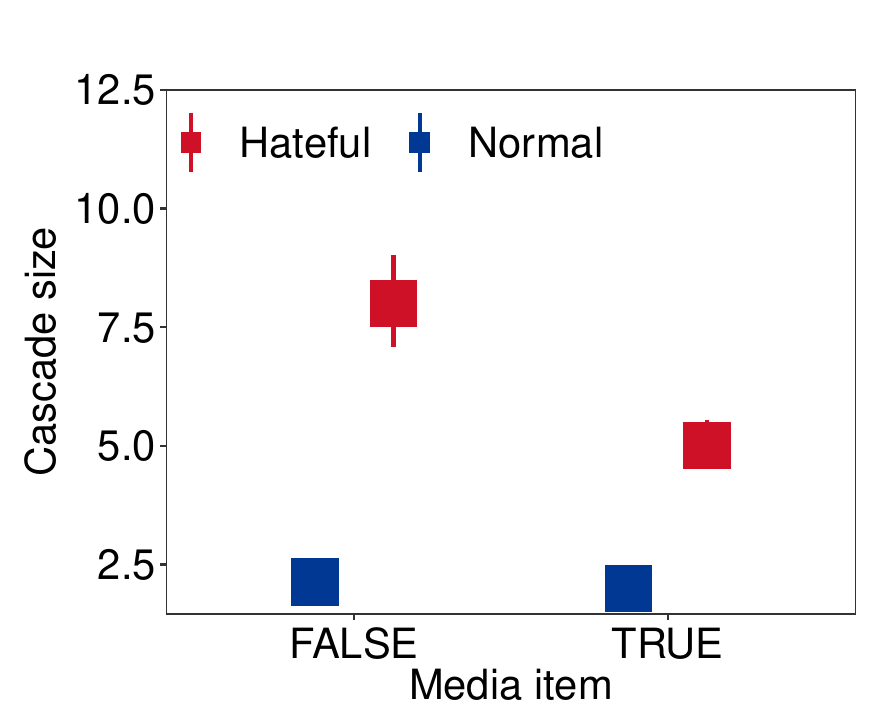}\\
\end{tabular}%
} \hfill
\subfloat[Cascade lifetime]{\label{fig:marginal_effects_liftime_2}
\begin{tabular}[b]{@{}cccc@{}}
\includegraphics[width=.25\linewidth]{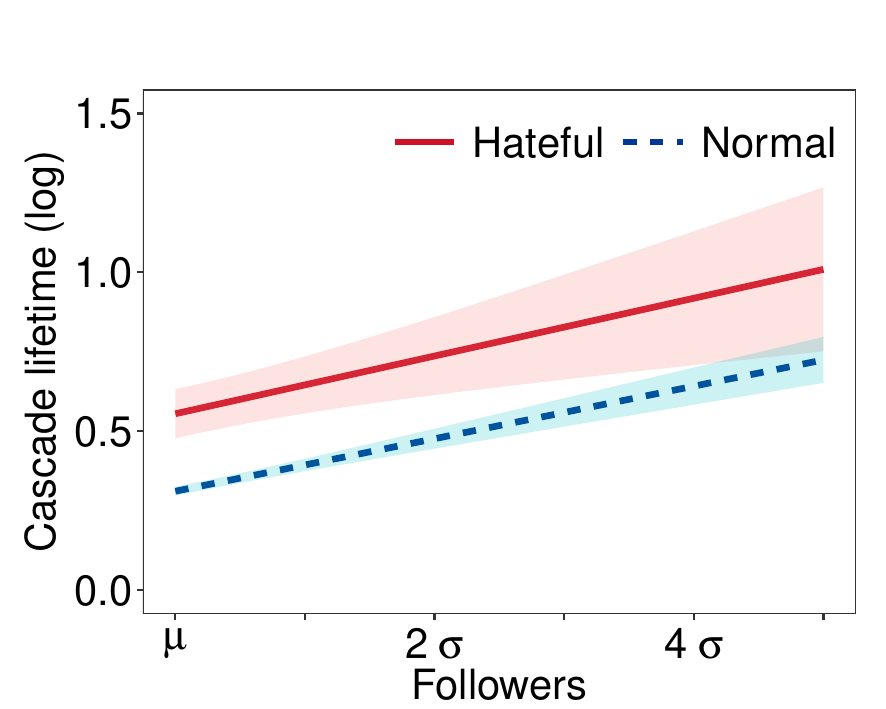}&
\includegraphics[width=.25\linewidth]{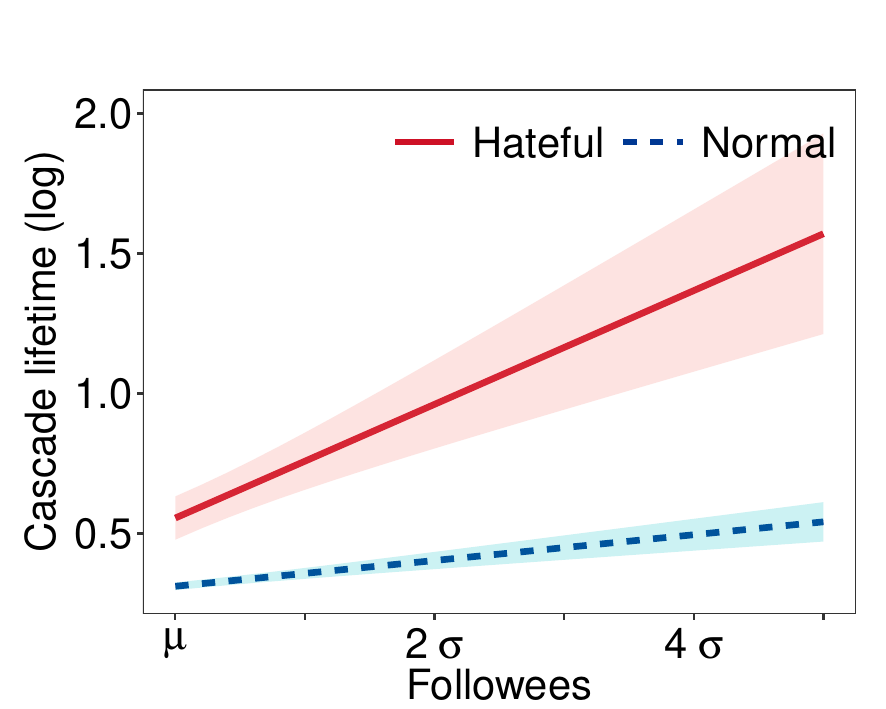}&
\includegraphics[width=.25\linewidth]{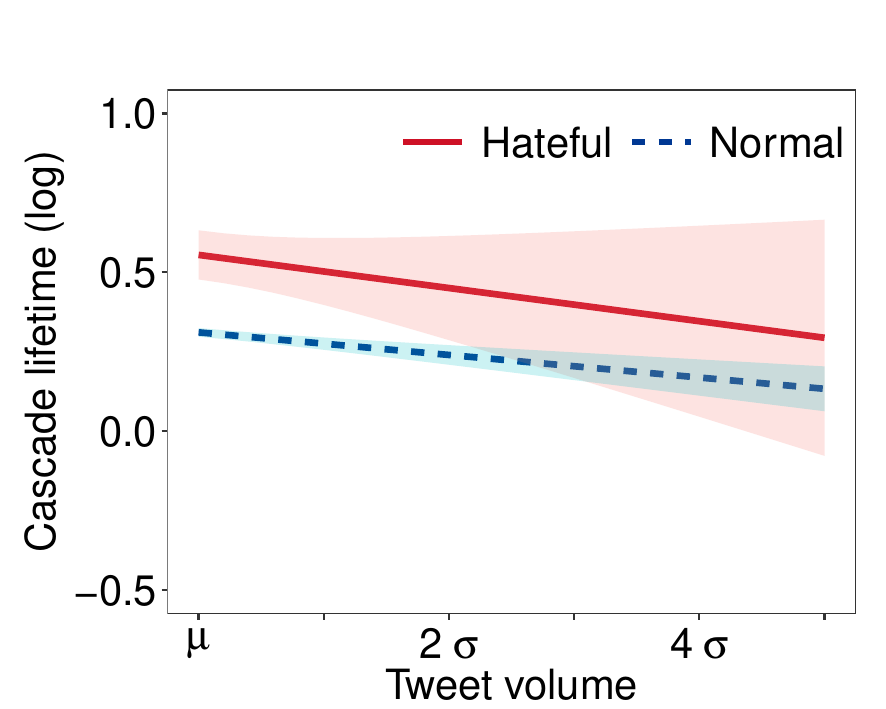}&
\includegraphics[width=.25\linewidth]{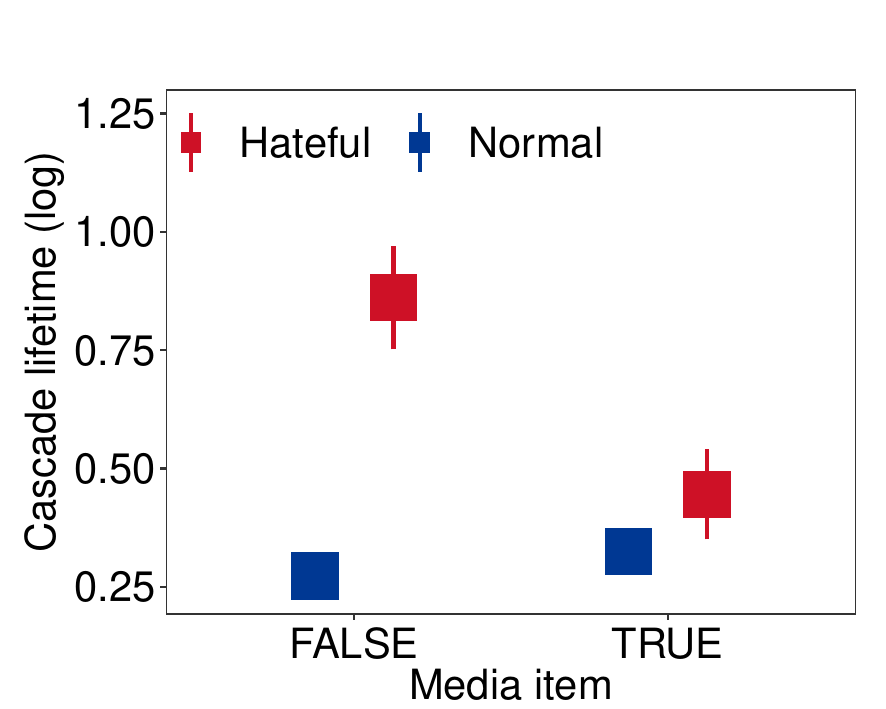} \\
\end{tabular}%
} \hfill
\subfloat[Structural virality]{\label{fig:marginal_effects_sv_2}
\begin{tabular}[b]{@{}cccc@{}}
\includegraphics[width=.25\linewidth]{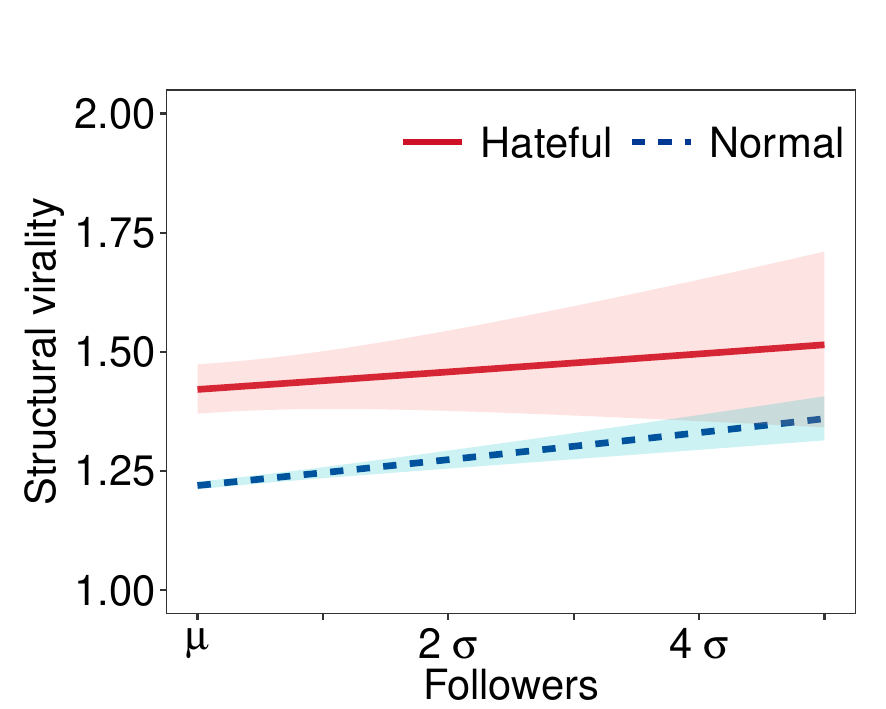}&%
\includegraphics[width=.25\linewidth]{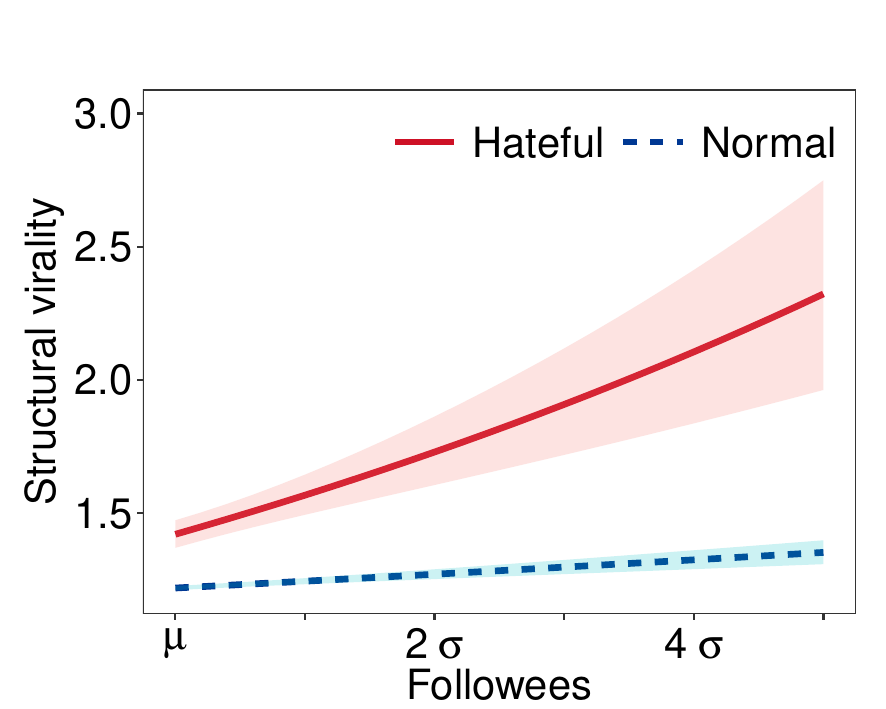}&
\includegraphics[width=.25\linewidth]{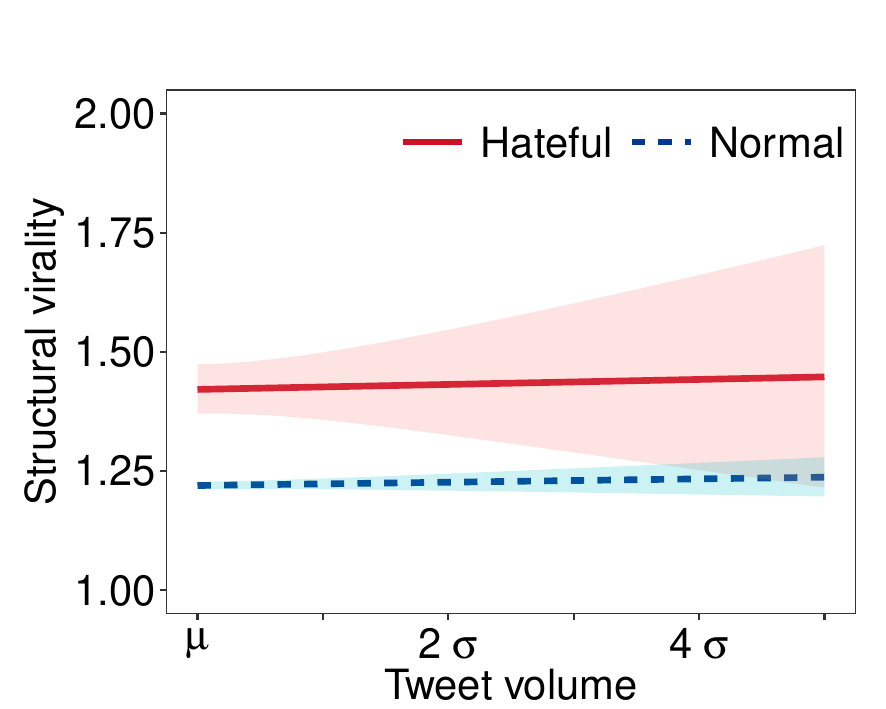}&
\includegraphics[width=.25\linewidth]{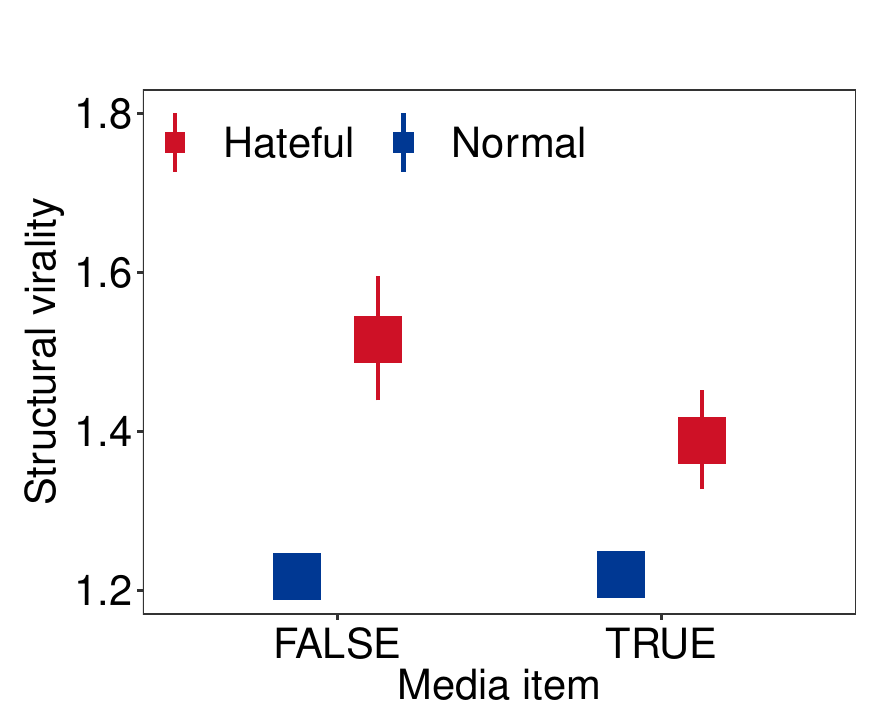} \\
\end{tabular}%
}
\addtolength{\tabcolsep}{7pt} 
\caption{Predicted marginal effects of the key determinants on (a)~cascade size, (b)~cascade lifetime, and (c)~structural virality. Shown is the mean estimate with \SI{95}{\percent} confidence intervals for hateful content (in \textcolor{BrickRed}{red}) and for normal content (in \textcolor{NavyBlue}{blue}). Continuous variables (here followers, followees, and tweet volume) were $\bm{z}$-standardized. Thus, their values on the $x$-axis represent the mean ($\bm{\mu}$) and standard deviations ($\bm{\sigma}$).
}
\label{fig:marginal_effects_2}
\end{figure*}

The predicted marginal effects for the remaining explanatory variables are shown in \Cref{fig:marginal_effects_2}. Account age is excluded as the corresponding coefficient estimates are not significant for cascade lifetime and structural virality as DVs.

%
%

\end{document}